\documentclass[12pt,preprint]{aastex}

\shorttitle{Molecular ion abundances in protoplanetary disks}
\shortauthors{Aikawa et al.}

\begin{document}


\title{Analytical Formulas of Molecular Ion Abundances and N$_2$H$^+$ Ring in Protoplanetary Disks}


\author{Yuri Aikawa}
\affil{Center for Computational Sciences, University of Tsukuba,
    1-1-1 Tennoudai, Tsukuba 305-8577, Japan}
\email{aikawa@ccs.tsukuba.ac.jp}

\author{Kenji Furuya}
\affil{Leiden Observatory, Leiden University, P.O. Box 9513, 2300 RA Leiden, The Netherlands}

\author{Hideko Nomura}
\affil{Department of Earth and Planetary Science, Tokyo Institute of Technology, 2-12-1
Ookayama, Meguro-ku, 152-8551 Tokyo, Japan}

\and

\author{Chunhua Qi}
\affil{Harvard-Smithsonian Center for Astrophysics, Cambridge,
MA 02138, USA.}

\begin{abstract}

We investigate the chemistry of ion molecules in protoplanetary disks, motivated by the detection of N$_2$H$^+$
ring around TW Hya. While the ring inner radius coincides with the CO snow line, it is not apparent why
N$_2$H$^+$ is abundant outside the CO snow line in spite of the similar sublimation temperatures of CO and N$_2$.
Using the full gas-grain network model, we reproduced the N$_2$H$^+$ ring in a disk model with millimeter grains.
The chemical conversion of CO and N$_2$ to less volatile species (sink effect hereinafter) is found to affect the
N$_2$H$^+$ distribution. Since the efficiency of the sink depends on various parameters such as activation barriers
of grain surface reactions, which are not well constrained, we also constructed the no-sink model; the total
(gas and ice) CO and N$_2$ abundances are set constant, and their gaseous abundances are given by the balance
between adsorption and desorption. Abundances of molecular ions in the no-sink model are calculated by analytical
formulas, which are derived by analyzing the full-network model. The N$_2$H$^+$ ring is reproduced by
the no-sink model, as well. The 2D (R-Z) distribution of N$_2$H$^+$, however, is different among the full-network
model and no-sink model.
The column density of N$_2$H$^+$ in the no-sink model
depends sensitively on the desorption rate of CO and N$_2$, and the flux of cosmic ray.
We also found that N$_2$H$^+$ abundance can peak at the temperature slightly below the CO sublimation,
even if the desorption energies of CO and N$_2$ are the same.

\end{abstract}





\section{Introduction}
A ring of N$_2$H$^+$ emission was found in the disk around TW Hya using ALMA (Atacama Large Millimeter
and submillimeter Array) \citep{qi13}. N$_2$H$^+$ is considered to be a useful probe of CO snow line, because
it is destroyed by the proton transfer to CO
\begin{equation}
{\rm N_2H^+ + CO \rightarrow  HCO^+ + N_2} \label{n2hp_co}.
\end{equation}
The inner radius of N$_2$H$^+$ ring is indeed consistent
with the CO snow line predicted from a disk model of TW Hya. The anti-correlation of N$_2$H$^+$ and CO
emission is also often observed in prestellar cores \citep[e.g.][]{tafalla04}, which strengthen the above statement.

The bright emission of N$_2$H$^+$ outside the CO snow line is, however, puzzling. Laboratory experiments show that
the sublimation temperatures of CO and N$_2$ are similar \citep{collings04,oberg05}. Since N$_2$H$^+$ is formed by protonation
of N$_2$, N$_2$H$^+$ should also be depleted outside the CO snow line. In the case of prestellar cores, there are two
possible explanations for the survival of N$_2$H$^+$ in the CO depleted region. Firstly, destruction of N$_2$H$^+$ is temporally
suppressed by the freeze-out of its major reactant, CO. Secondly, the slow formation of N$_2$ from N atom in molecular clouds
helps the temporal survival of N$_2$ \citep{aikawa01,maret06,bergin07}.
The first mechanism could work in the disk as well, in a narrow range of temperatures where the positive effect of depletion
on N$_2$H$^+$ abundance (i.e. decrease of CO) wins against the negative effect (i.e. depletion of N$_2$).
The second mechanism, on the other hand, seems irrelevant for disks. In protoplanetary disks, in which the gas density
is much higher than molecular clouds, it is not likely that N atom is more abundant than N$_2$ except for
the photodissociation layer at disk surface.

In chemistry models of disks, it is often found that CO is depleted even in regions
warmer than its sublimation temperature ($\sim 20$ K) via conversion to less volatile species
\citep{aikawa97,bergin14,furuya14}. Since the conversion works as a sink in the chemical reaction network of gaseous
species, we call it the sink effect in the present work.
\cite{favre13} showed, for the first time, observational evidences for such CO depletion
towards TW Hya. \cite{furuya14} showed that N$_2$ is also subject to the sink effect; it is converted to NH$_3$ via the
gas-phase reactions of 
\begin{eqnarray}
{\rm N}_2 + {\rm H}_3^+ \rightarrow {\rm N}_2{\rm H}^+ + {\rm H}_2 \label{N2_H3p} \\
{\rm N}_2{\rm H}^+ + {\rm e} \rightarrow {\rm NH} + {\rm N} \label{N2Hp_recomb} 
\end{eqnarray}
and subsequent hydrogenations of NH on grain surfaces. It should be noted that this conversion is not efficient when
CO is abundant, because N$_2$H$^+$ mainly reacts with CO to reform N$_2$. 
The longer timescale required for N$_2$ depletion could explain the N$_2$H$^+$ ring.
But then the lifetime of the ring should be similar to the difference in the depletion timescales
of CO and N$_2$, which is rather short $\lesssim 10^5$ yr, if the ionization rate is
$\sim 5\times 10^{-17}$ s$^{-1}$ \citep{furuya14}.
Since N$_2$ is considered to be a major nitrogen carrier in protoplanetary disks, but cannot be directly
observed, it is important to constrain its distribution based on N$_2$H$^+$ observation.
The formation mechanism of N$_2$H$^+$ ring is thus worth investigation via theoretical calculations.

The abundance of N$_2$H$^+$ could also be a probe of ionization degree in the regions of CO depletion.
The ionization degree depends on the gas density and ionization rate,
which is an important parameter for both chemical and physical evolutions of the disk.
Since ionizations trigger the chemical reactions, as in molecular clouds, the timescale of chemical evolution
depends on the ionization rate \citep{aikawa97}. 
The ionization
degree determines the coupling with magnetic fields, e.g. which disk region is subject to magneto-rotational
instability \citep{balbus91}. The major ionization sources are X-ray from the central star, cosmic ray,
and decay of radioactive nuclei \citep{glassgold97,umebayashi88,cleeves13}. The ionization rate thus depends
on the flux and hardness of X-ray radiation and the abundances of radioactive nuclei. In addition, the stellar winds
and/or magnetic fields of the star-disk system could prevent the penetration of the cosmic ray to the disk.
Since these parameters are unknown and could vary among objects, the ionization degree should be probed via observations.
The observation of ionization degree is principally possible using major ion molecules. But a quantitative estimate
of the ionization degree from the observational data is not straightforward, because the gas density,
major ion molecules, and their abundances change spatially within the disk \citep[e.g][]{aikawa02, qi08, cleeves14}.

In this work, we investigate the spatial distribution of N$_2$H$^+$ and other major molecular ions in protoplanetary
disks. Firstly, we calculate the full chemical network model, which includes gas-phase and grain-surface reactions.
The radial distribution of N$_2$H$^+$ column density has a peak around the radius of CO sublimation temperature;
i.e. the N$_2$H$^+$ ring is
reproduced. We found that the sink effect on CO significantly affects the distribution of N$_2$H$^+$. 
By analyzing the full chemical network, we also found that the abundances of major molecular ions, H$_3^+$, HCO$^+$ and
N$_2$H$^+$ can be described by analytical functions of gas density, temperature, ionization rate, and abundances of CO and N$_2$.
Such formulas are useful in deriving the ionization degree from the observations of molecular ions, combined with
the dust continuum and CO observations to constrain the gas density and CO abundance.

While the sink effect plays an important role in determining the N$_2$H$^+$ distribution in the full network model,
the efficiency of the sink depends on various parameters which are not well constrained yet.
For example, the conversion timescale of CO and N$_2$ to less volatile species depends on ionization rate and activation barriers
of the reactions. The sink effect could be less significant, if the disk is turbulent and the diffusion timescale is shorter
than the conversion timescale. Since there are various chemical paths for the conversion,
it is difficult to directly control the efficiency of sink in the full-network model. In order to investigate the abundances
of N$_2$H$^+$ and other molecular ions in the limit of no sink, we construct "no-sink" model; we assume that
the total (gas and ice) abundances of CO and N$_2$ are constant, and that their gas-phase abundances are given by the balance
between adsorption and desorption. The abundances of molecular ions are calculated using the analytical formulas.

The plan of this paper is as follows. In \S 2, we describe our disk model and chemical model. 
The molecular distributions in the full-network model
are presented in section \S 3. In \S 4, we derive analytical formulas for abundances of electron, 
H$_3^+$, HCO$^+$ and N$_2$H$^+$. The formulas are used to calculate their abundances in the
disk model, which are compared with the results of full-network calculation. Section 5 presents the
results of no-sink model. Since the no-sink model is analytical,
we can easily investigate the dependence of N$_2$H$^+$ column density on desorption rate (i.e. sublimation temperature)
of CO and N$_2$. We also investigate how much the N$_2$H$^+$
column density is reduced, if the penetration of cosmic rays to the disk is hampered by the stellar wind and/or
magnetic fields. We summarize our results and conclusions in \S 6.

\section{Models}
We adopt the same disk model as in \cite{furuya14}. In addition to the disk model with small grains,
which is assumed in \cite{furuya14}, we also investigate a disk model in which the dust grains have
grown up to a radius of 1 mm. The turbulent mixing is not explicitly included in the present work.
The model of full chemical network is also the same as in \cite{furuya14}, but includes some updates.
The models are briefly described in the following.

\subsection{Disk Model}
\begin{figure}
\epsscale{.80}
\plotone{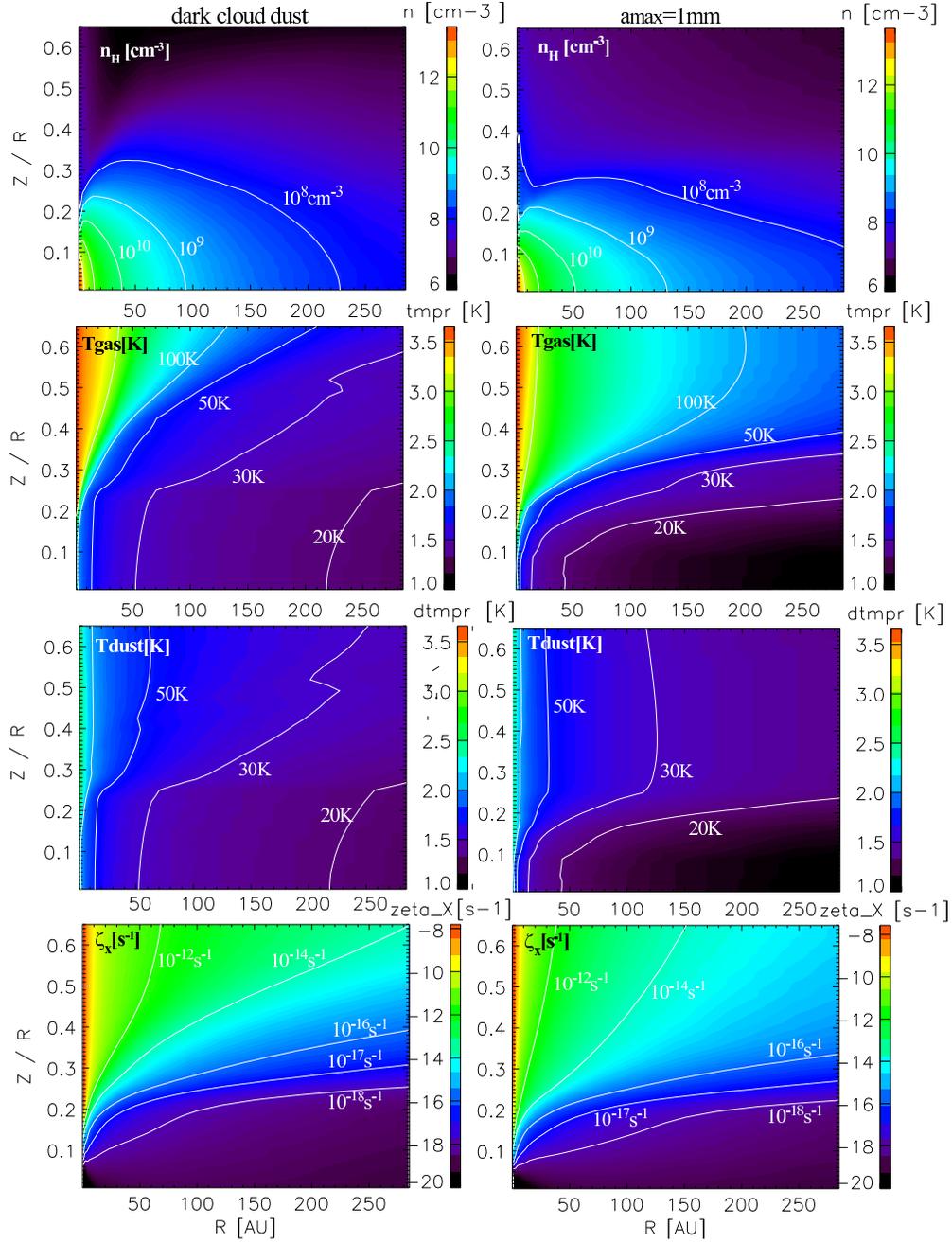}
\caption{Distribution of the density of hydrogen nuclei $n_{\rm H}$(top), gas temperature, dust temperature, and
ionization rate by X-rays $\zeta_{\rm X}$ (bottom) in the disk model with dark cloud dust (left panels) and millimeter grains (right panels).
\label{disk_model}}
\end{figure}

Since we aim to understand the mechanism to form N$_2$H$^+$ ring, rather than to investigate the disk structure of a specific
object (i.e. TW Hya), we adopt a steady, axisymmetric Keplerian disk around a T Tauri star.
The stellar mass, radius and effective temperature are $M_* = 0.5 M_{\odot}$, $R_*= 2 R_{\odot}$ and
$T_*= 4000$ K, respectively. The disk structure is given by solving the radiation transfer, thermal balance
of gas and dust, and hydrostatic equilibrium in the vertical direction in the disk. Basic equations and
calculation procedures are described in \cite{nomura07}.
We assume the stellar UV and X-ray luminosity of $10^{31}$ erg s$^{-1}$ and $10^{30}$ erg s$^{-1}$,
respectively. The cosmic ray ionization rate of H$_2$ is set to be $5\times 10^{-17}$ s$^{-1}$ \citep{dalgarno06},
while the ionization rate by the decay of radioactive nuclei is set to be $1\times 10^{-18}$ s$^{-1}$ \citep{umebayashi09}.
The dust-to-gas mass ratio is 0.01. We consider two disk models: one with the dark cloud dust and the other with millimeter-sized
grains \citep{aikawa06}.
The former assumes the dust properties of \cite{weingartner01} ($R_{\rm v}=5.5, b_{\rm c}=3\times 10^{-5}$,
case B); while the silicate grains have a rather steep size distribution with the maximum radius of $\sim 0.2$ $\mu$m,
the carbonaceous grains have PAH-like properties in the small-size limit and graphite-like properties
at larger sizes. The maximum size of carbonaceous grain is $\sim 10$ $\mu$m.
In the latter model, we assume the power-law size distribution of dust grains
$dn(a)/da\propto a^{-3.5}$, where $a$ is the
grain radius, referring to the ISM dust model of \cite{mrn77}, but the minimum and maximum sizes are set to be 0.01 $\mu$m
and 1 mm, respectively. It would be more appropriate for T Tauri disks than the dark cloud dust model, since the
grain growth is indicated by the disk observations \citep[e.g.][]{williams11}.
The dust opacities for the two models are calculated using Mie theory.
The gas temperature, dust temperature, and density distributions in the disk are
calculated self-consistently, by considering various heating and
cooling mechanisms. Figure \ref{disk_model} shows the distribution of gas density, gas temperature, dust temperature, and ionization
rate by X-ray in our models. Cosmic ray ionization dominates in the midplane where the X-ray ionization rate is
$\le 5\times 10^{-17}$ s$^{-1}$.
The temperature in the model with millimeter grains is lower than that in the dark cloud dust model due to the lower dust opacity
at a given disk height ($Z$) to receive the radiation from the central star. Column density in these disk models are
determined by assuming a steady state disk structure with constant viscosity and accretion rate (although we
consider turbulent diffusion and/or radial accretion only implicitly in the chemical model), so that the masses of the
two disks are slightly different: $1\times 10^{-2} M_{\odot}$ for the dark cloud dust model and $1.7\times 10^{-2} M_{\odot}$ for
the millimeter grain model.

\subsection{Chemical Model: Full Network}

Our chemical network is based on \cite{garrod06}. We added photo-ionization, photo-dissociation
and photo-desorption by UV radiation from the central star, self-shielding of H$_2$, CO and N$_2$
\citep{furuya13, li13}, X-ray chemistry and charge balance 
of dust grains. Although our model includes Deuteration and ortho/para states of several species
such as H$_2$ and H$_3^+$ (Hincelin et al. in prep, Furuya et al. in prep), we present only the molecular
abundances, i.e. the sum of isotopomers
and o/p states. The D/H ratio and o/p ratios will be presented in forthcoming papers.
Our model consists of two phases, gas phase and ice mantle; i.e. we do not
discriminate layers of ice mantles, unless otherwise stated.
Desorption energies ($E_{\rm des}$) of assorted species are listed in Table 1 in \cite{furuya14}.
Desorption energies of atomic hydrogen, CO and N$_2$ are  set to be 600 K, 1150 K, and 1000 K,
respectively; they are the values on water ice substrates \citep{alhalabi07,garrod06}.
The sublimation temperature of CO and N$_2$ are then $\sim 23$ K and $\sim 19$ K, when the
gas density is $10^6$ cm$^{-3}$. We investigate the dependence of N$_2$H$^+$ abundance on the
desorption energies of CO and N$_2$ in \S 3 and \S 5.


Adsorbed species on grains migrate via thermal hopping and react with each other when they meet.
We adopt the modified rate of \cite{caselli98,caselli02} for grain surface reactions of H atom.  
The adsorption rate of gaseous species onto grain surfaces and grain surface reaction rates (e.g.
if the rate is limited by the accretion of gaseous particle) depend on the size distribution of grains.
Ideally, the grain size distribution should be taken into account \citep{acharyya11}.
Most chemical models of disks and molecular clouds, however, assume a single size of 0.1 $\mu$m, for
simplicity, which we follow in the present work.
The rate coefficients of gas-dust interactions (e.g. adsorption) are basically proportional to the total
grain surface area. The assumption of a single grain size in the chemical model is thus a reasonable
approximation, as long as the total surface area of grains are consistent with that in the physical disk model.
The total surface area of the 0.1 $\mu$m dust model agrees with that of our dark cloud dust model within a factor of
a few. In the chemical model for the disk model with millimeter grains,
we adopt the same uniform grain size (0.1 $\mu$m), but decrease the dust-gas
ratio by one order of magnitude; according to the power-law size distribution,
the number of small grains, which dominate in the grain surface area, is decreased compared with the dark cloud dust model
by an order of magnitude.
\citep{aikawa06}.
One caveat for this single-size approximation is that we may underestimate the rate of grain surface recombination.
In the dense regions, such as disk midplane at small radii, recombinations are more efficient on grain surfaces than in
the gas phase. Due to the Coulomb focusing, the cross section of grain-surface recombination is much larger than the
geometrical cross section of dust grains. The rate of grain surface recombination thus cannot be scaled by the
total grain surface area.

The photodissociation rates are calculated by convolving the attenuated stellar and interstellar
UV spectrum and wavelength-dependent photodissociation cross sections at each position in the disk \citep{aikawa02, vandishoeck06}.
UV radiation induced by X-ray and cosmic ray \citep{gredel89} is also taken into account.
For ice mantle species, we assume that only the uppermost layers can be dissociated; i.e.
while the UV radiation can penetrate into deep layers of ice mantle, we assume that
the photo products in deeper layers recombine immediately. The effective rate of photodissociation is
thus reduced. Considering the fluffiness and pores on
the grain surfaces, we let the uppermost two layers, rather than one, to be dissociated.

We take into account three non-thermal desorption processes: photodesorption, stochastic heating
by cosmic-rays, and reactive desorption.
We adopt the photodesorption yields per incident FUV photon
derived from the laboratory experiment for H$_2$O, CO$_2$, CO, O$_2$ , and N$_2$
\citep{oberg09a, oberg09b,fayolle11, fayolle13}. A yield is set to $10^{-3}$ for other species.

Initial molecular abundance in the disk is given by calculating the molecular evolution
in a star-forming core model of \citet{masunaga00} \citep[see also][]{masunaga98, aikawa08}.
The initial abundances of assorted molecules and the elemental abundances in our model
are listed in Table \ref{tab_init_abun}.
Major carriers of oxygen and carbon are H$_2$O and CO, while the major N-bearing species are
NH$_3$ and N$_2$. We adopt the low-metal abundance; i.e. the abundances of metals such as Mg
and Si are about two orders of magnitude lower than observed in diffuse clouds.

We calculate the chemical reaction network (i.e. rate equations) as an initial value problem at each position
in the disk. As we will see in \S 3, the abundances of electron, HCO$^+$, N$_2$H$^+$, and H$_3^+$
reach the steady state, which are
determined by the ionization rate, gas density, temperature, and the abundances of CO and N$_2$, in a short timescale.
On the other hand, CO and N$_2$ decrease slowly with time mainly due to the sink effect.
Vertical diffusion and radial accretion, which are not explicitly included in
the present work, could suppress or slow-down the sink effect \citep{furuya13, furuya14}. Therefore we
present molecular abundances at an early time $1\times 10^5$ yr, as well as at the typical timescale of
T Tauri stars $\sim 10^6$ yr.

\subsection{Chemical Model: No-Sink Model}

In the full-network model, we will see that the distribution of N$_2$H$^+$ is significantly 
affected by the CO depletion via sink effect. Efficiency of the sink effect, however, depends on various parameters
such as CO$_2$ formation (CO + OH) rate on dust grains, initial CO abundance, ionization rate,
and turbulent mixing \citep{bergin14,furuya14}.
A strong vertical turbulence, for example, tends to smooth out the molecular abundances, so that
the local CO abundance minima due to the sink could be less significant.
Although the observation of TW Hya indicates the CO depletion via sink effect \citep{favre13},
the spatial distribution of CO abundance is not well constrained yet. It is therefore useful to calculate the distribution
of N$_2$H$^+$ in a model without the sink effect. 

In the no-sink model, we assume that the sum of gas-phase and ice-mantle abundances of CO is equal to its canonical abundance:
i.e. $1\times 10^{-4}$ relative to the hydrogen nuclei. The gas-phase abundance of CO is given by a simple balance between
adsorption and desorption;
\begin{eqnarray}
&&\frac{n_{\rm COgas}}{n_{\rm COice}}=\frac{\nu \exp (-\frac{E_{\rm des}(CO)}{kT})+\nu\tau_{\rm CR}C_{\rm Fe}
\exp (-\frac{E_{\rm des}(CO)}{kT_{\rm max}}) } {S \pi a^2 n_{\rm{dust}} v_{\rm th} } \label{2phase} \\
&&n_{\rm{COgas}}+n_{\rm{COice}}=10^{-4} n_{\rm{H}} \label{CO_analytical}, 
\end{eqnarray}
where $n_{\rm COgas}$ and $n_{\rm COice}$ are number densities of CO in the gas phase and ice mantle.
While the first term in the numerator represents the thermal desorption, the second term represents the non-thermal desorption.
Although our full network model includes various mechanisms of non-thermal desorption, here we consider only the
stochastic heating by cosmic-rays for simplicity. It is the effective desorption mechanism in the
cold midplane for species with relatively low desorption energies \citep{hase93}. 
 The frequency of CO oscillation on grain surface $\nu$ is set to be $10^{12}$ s$^{-1}$.
When a cosmic-ray particle hits a dust grain, the grain is heated temporally for $\tau_{\rm CR}\sim 10^{-5}$ sec.
The peak temperature of the temporal heating is set to be $T_{\rm max}=70$ K.
The rate for a grain to encounter Fe ion particle, which is the most efficient in dust heating among cosmic-ray
particles, is $C_{\rm{Fe}}=3\times 10^{-14}$ s$^{-1}$.
The denominator represents the sticking rate of gaseous CO onto grain surfaces. The sticking probability on collision is
set to be $S=1.0$. The grain size $a$ is 0.1 $\mu$m (see \S 2.2), and $v_{\rm th}$ is the thermal velocity of CO particle.
The abundance of gaseous N$_2$ is formulated similarly, with the total N$_2$ abundance set to be $4.5\times 10^{-6}$.
It should be noted that this desorption rate by the cosmic-ray heating is a rough estimate. In reality and in our physical
disk models, the grains actually have a size distribution, and the parameters such as $\tau_{\rm CR}$ and $T_{\rm max}$
depend on the grain size. \citet{leger85}, however, showed that the desorption rate via cosmic-ray heating does not sensitively
depend on the grain size as long as the grains are small $\lesssim 0.2$ $\mu$m. Since the small grains contribute most to the
total surface area of grains, the desorption rate obtained here would be reasonable.

In section \S 4, we analyze the results of full network model to find out that the abundances of electron,
H$_3^+$, HCO$^+$, and N$_2$H$^+$ can be well described by analytical formulas, which are the functions of density,
temperature, ionization rate, and abundances of CO and N$_2$. We use these analytical formulas to obtain
the molecular ion abundances in the no-sink model.
A combination of the analytical formulas of molecular ions and equilibrium abundances of gaseous CO and N$_2$ (eq.
\ref{2phase} and \ref{CO_analytical}) makes it very easy to investigate
the dependence of N$_2$H$^+$ abundance on various parameters such as desorption energies of CO and N$_2$ and ionization
rate in the disk.

\section{Results: Full Network}
\subsection{Disk with dark cloud dust}
Figure \ref{ISM_chem} shows the distributions of CO, HCO$^+$, N$_2$, N$_2$H$^+$, H$_3^+$ and electron abundances at
the time of $1\times 10^5$ and $9.3 \times 10^5$ yr in the model with dark cloud dust. 
N$_2$H$^+$ exists mostly in the upper layers of the disk.
H$_3^+$ is more abundant than HCO$^+$ in the disk surface due
to a relatively high abundance ratio of electron to CO (see \S 4.4). N$_2$H$^+$ is thus kept
abundant there, via N$_2$ + H$_3^+$, in spite of the destruction via the reaction (\ref{n2hp_co}).

The dashed lines in the panels of CO and N$_2$ depict the position where the gas-phase abundance and ice abundance
become equal in the adsorption-desorption equilibrium:
\begin{eqnarray}
&&\frac{n_{\rm gas}}{n_{\rm ice}}=\frac{\nu \exp (-\frac{E_{\rm des}}{kT})+\nu\tau_{\rm CR}C_{\rm Fe}
\exp (-\frac{E_{\rm des}}{kT_{\rm max}}) } {S \pi a^2 n_{\rm{dust}} v_{\rm th} }=1, \label{snowline}
\end{eqnarray}
where $n_{\rm gas}$ and $n_{\rm ice}$ are number densities of CO (or N$_2$) in the gas phase and ice mantle.
It defines the "snow surface" of volatile species in the disk.
We define the CO and N$_2$ "snow line" as the radius at which the equation (\ref{snowline}) is satisfied in the midplane.

In Figure \ref{ISM_chem}, we can see that both CO and N$_2$ are depleted in the midplane even inside their snow lines,
especially at the later stage, $t=9.3 \times 10^5$ yr;  CO is converted to CO$_2$ ice, CH$_3$OH ice and hydrocarbons
such as CH$_4$ ice and C$_2$H$_6$ ice, while N$_2$ is converted to NH$_3$ ice. N$_2$H$^+$ is not
abundant in the CO depleted region, since its mother molecule, N$_2$, is depleted there as well.
At $t=1\times 10^5$ yr, the sink effect is still moderate, and N$_2$H$^+$ abundance in the midplane has a local peak
at $\sim 200$ AU. N$_2$ is depleted in the outer radius, while N$_2$H$^+$ is
destroyed by CO in the inner radius.

\subsection{Disk with millimeter grains}
Figure \ref{mm_chem} shows the distributions of CO, HCO$^+$, N$_2$, N$_2$H$^+$, H$_3^+$ and electron abundances in
the disk model with millimeter grains.
The midplane temperature is warmer than
20 K and 30 K inside $\sim 40$ AU and $\sim 10$ AU, respectively. Although the sink effect is less significant
due to the smaller total surface area of dust grains than in the model with dark cloud dust, CO is converted to
CO$_2$ ice via the grain-surface reaction of CO + OH, and is depleted from the gas phase even in the intermediate
($Z/R\sim 0.2$) layers above the dashed line, where the dust temperature is higher than the sublimation temperature of CO.
In the midplane region with $T\lesssim 20$ K, on the other hand, major carbon reservoirs are
CO ice, CO$_2$ ice and CH$_3$OH ice.

While the spatial distributions of CO and N$_2$ are similar in the dark cloud dust model, they are significantly
different in the millimeter grain model; N$_2$ is abundant in a layer at $Z/R\sim 0.2$, where CO is depleted.
In regions closer to the midplane, N$_2$ ice and NH$_3$ ice are the dominant N-bearing species.
The conversion of N$_2$ to NH$_3$ ice proceeds via the gas-phase reactions of (\ref{N2_H3p}) and (\ref{N2Hp_recomb}),
and subsequent hydrogenation of NH on grain surfaces. At $Z/R \sim 0.2$, the photodissociation of NH
(NH $\rightarrow$ N + H) is effective, which prevent the conversion of N$_2$ to NH$_3$ ice. The product of
photodissociation, N atom, is converted back to N$_2$ via the reaction of N + NO.
In other words, a deeper penetration of UV radiation in the millimeter grain model is a key to suppress the sink effect
on N$_2$.


Another key difference between the models with dark cloud dust and millimeter grains is that H$_3^+$ is not the dominant
ion in the disk surface in the latter model. Since the UV shielding via dust grains is less effective
in the millimeter grain model, UV radiation ionizes atoms, such as C and S, to make electrons abundant.
Dissociative recombination of a molecular ion with an electron is much more efficient than the radiative
recombination of an atomic ion. Thus the photoionization of atoms results in depletion of H$_3^+$ and N$_2$H$^+$
in the disk surface (e.g. $Z/R \gtrsim 0.3$).
In lower layers, the distribution of N$_2$H$^+$ is basically similar to that of N$_2$,
except for the midplane in inner radius where CO is abundant (i.e. $R<$ a few tens of AU at $t=1\times 10^5$ AU).

It should be noted that the layer with abundant N$_2$ and CO depletion (and thus abundant N$_2$H$^+$)
at $Z/R\sim 0.2$ is formed via the combination of sink effect on CO and photodissociation of NH,
rather than via the difference in sublimation
temperature of CO and N$_2$. We performed a calculation of the same disk model but
setting the both desorption
energies of CO and N$_2$ to be 1000 K. Distributions of CO, HCO$^+$, N$_2$ and N$_2$H$^+$ are shown in
Figure \ref{CO1000_chem}. 
The resultant distribution of molecules are basically the same as in Figure \ref{mm_chem}.

\subsection{Column densities}
Although we do not aim to construct a best-fit model for a specific object, it is useful to briefly compare our models with
observations to see which model reproduces the N$_2$H$^+$ ring better. Since the radiative transfer calculation is out of the scope
of the present work, we compare the column density of N$_2$H$^+$ in our full-network models with the estimated values in TW Hya.

In the N$_2$H$^+$ observation by \citet{qi13}, the 1-sigma detection limit corresponds to the N$_2$H$^+$ column density of
$2\times 10^{11}$ cm$^{-2}$. \citet{qi13} constructed disk models to fit their observational data. In the models that can
reasonably fit the observational data, the N$_2$H$^+$ column density at the inner edge of the ring ranges from $4\times 10^{12}$
to $2\times 10^{15}$ cm$^{-2}$. They also found that the column density contrast at the inner edge of the N$_2$H$^+$ ring
is at least 20, and could be larger. 

We also calculated the optical depth of N$_2$H$^+$ ($J=4-3$) line for a slab of H$_2$ and N$_2$H$^+$ gas using
the Radex code \citep{vandertak07}. Assuming the H$_2$ density of $10^8$ cm$^{-3}$ and  line width of 0.15 km s$^{-1}$,
the N$_2$H$^+$ column density of $10^{12}$ cm$^{-2}$ corresponds to the optical depth of
$\tau=0.60$ and 0.49 for the gas temperature of 17 K and 30 K, respectively. In order to reproduce N$_2$H$^+$ ring,
the column density of N$_2$H$^+$ should be at least higher than $\sim 10^{12}$ cm$^{-2}$, which is consistent with \citet{qi13}.

Radial distributions of column densities of CO, HCO$^+$, N$_2$ and N$_2$H$^+$ in the dark cloud dust model are plotted in
Figure \ref{column}(a) and (c). The horizontal axis, $R$, is linear in (a),
while it is logarithmic to highlight the inner radius in (c).
The distribution of N$_2$H$^+$ column density is rather flat. At $t=1\times 10^5$ yr, the N$_2$H$^+$ abundance has a peak
in the midplane at $R\sim 200$ AU, which is outside the CO snow line. But N$_2$H$^+$ column density does not
sharply drop inwards at the CO snow line ($\sim 125$ AU), because N$_2$H$^+$ is abundant in the disk surface even in the
inner radius.
At $9.3\times 10^5$ yr, the N$_2$H$^+$ is abundant only
in the disk surface, and the N$_2$H$^+$ column density shows a peak at $\sim 30$ AU, which is inside the CO snow line.

The column densities of molecules in the millimeter grain model are shown in Figure \ref{column} (b) and (d).
At $t=1\times 10^5$ yr, the N$_2$H$^+$ column density has a peak at the radius of $\sim 40$ AU, and significantly decreases
inwards at $\sim 20$ AU, which corresponds to the CO snow line in the disk model with millimeter grains.
This model is thus in better agreement with the observation of TW Hya than the dark cloud dust model.
At $t=9.3 \times 10^5$ yr, the radius of
N$_2$H$^+$ column density peak is shifted inwards due to the sink effect on CO and N$_2$.


\section{Analytical Solution for Molecular Ion Abundances}
In the molecular layer, the major ions are HCO$^+$, H$_3^+$ and N$_2$H$^+$.
In this section we derive analytical formulas for their abundances.
The analysis helps us to better understand the spatial distributions of these species (Figure \ref{ISM_chem}
and \ref{mm_chem}). We will show that we can calculate HCO$^+$ and N$_2$H$^+$ abundances in a disk model,
if the distributions of gas density, temperature, ionization rate, and abundances of CO and N$_2$ are given.
The formulas are useful in constraining the ionization rate from the observation of CO, HCO$^+$ and N$_2$H$^+$.

\subsection{Electron}
Electron abundance, i.e. the ionization degree, is determined by the balance between ionization and recombination.
Let us first assume that HCO$^+$ is the dominant ion in the disk, for simplicity. 
A sequence of reactions starts with ionization of H$_2$ via cosmic-ray, X-ray, or decay of radioactive nuclei;
\begin{eqnarray}
&& \rm{H}_2  \rightarrow \rm{H}_2^+ + \rm{e}  \\
&& {\rm H}_2^+ + {\rm H}_2 \rightarrow {\rm H}_3^+ + {\rm H}  \label{h2p}\\
&& {\rm H}_3^+ + {\rm CO} \rightarrow {\rm HCO}^+ + {\rm H}_2  \\
&& \rm{HCO}^+ + {\rm e} \rightarrow \rm{H} + \rm{CO}. \label{rec_hcop}
\end{eqnarray}
We define the ionization degree $x$(e) as $n$(e)/$n_{\rm H}$ rather than $n$(e)/$n$(H$_2$); while the latter is
the usual definition, molecular abundances are relative to hydrogen nuclei in our numerical calculation.
Considering the neutrality, the ionization degree in molecular gas is given by
\begin{equation}
x({\rm e})=\frac{n({\rm e})}{n_{\rm H}}=\frac{n({\rm e})}{2n({\rm H}_2)}=\sqrt{\frac{\zeta}{2k_{\ref{rec_hcop}}n_{\rm H}}} \label{electron}.
\end{equation}
The ionization rate $\zeta$ is a sum of the rates by cosmic ray, X-ray (Figure \ref{disk_model}), and decay of radioactive nuclei.
The rate coefficients of relevant reactions are listed in Table \ref{tab_reactions} in Appendix A.

Now, it should be noted that the dominant ion varies within the disk.
Although the rate coefficients of the dissociative recombination of molecular ions are mostly of the same order
($\sim 10^{-7}$ cm$^3$ s$^{-1}$), the values vary slightly among molecular ions.
We thus adopt an iteration to evaluate the ionization degree.
Initially, we assume that the reaction (\ref{rec_hcop}) dominates in the recombination of electron in the gas phase,
and calculate the ionization degree, which is used to evaluate the abundances of molecular ions (see the following subsections).
Then we re-calculate the equation (\ref{electron}) by replacing $k_{\ref{rec_hcop}}$ with the average of the recombination rate
coefficients of molecular ions weighted by their abundances.

In the midplane at $R \lesssim$ a few 10 AU, the gas density is so high that the grain surface recombination
becomes more effective than the gas-phase recombination;
\begin{equation}
{\rm HCO^+ + G(-) \rightarrow  H + CO + G(0), ~ ~\label{hcop_g} }
\end{equation}
where G(-) and G(0) represent a negatively-charged grain and a neutral grain, respectively.
Then the ionization balance is described as
\begin{equation}
\zeta n({\rm H}_2)=k_{\ref{rec_hcop}}n({\rm HCO}^+)n({\rm e})+k_{\rm G}n({\rm G-})n({\rm HCO}^+)=k_{\ref{rec_hcop}}n^2({\rm e})+k_{\rm G}n({\rm G-})n({\rm e}),
\label{quad_e}
\end{equation}
where $k_{\rm G}$ is the rate coefficient of the grain surface recombination of HCO$^+$ \citep[see e.g.][]{umebayashi83}.
Assuming that most grains are negatively charged, which is valid when $n_{\rm H}/\zeta \lesssim 10^{30}$ cm$^{-3}$ s
\citep{umebayashi83}, the electron abundances is calculated by solving this quadratic function.
It should be noted that the equation (\ref{quad_e}) is equivalent to the equation (\ref{electron}),
when the gas-phase recombination is more effective than the grain-surface recombination. Thus we use the equation (\ref{quad_e})
rather than (\ref{electron}) in the calculation of ionization degree in the whole disk, and
$k_{\ref{rec_hcop}}$ is replaced by the weighted mean of the recombination rate coefficients in the iteration.

In deriving the equations (\ref{electron}) and (\ref{quad_e}), we have assumed that the molecular ions are the dominant charge carrier.
In the disk surface however, atomic ions such as C$^+$ and S$^+$ are produced via photoionization and dominate over molecular ions (\S 3.2).
In the transition layer from such an atomic-ion dominated (AID) layer to the molecular layer, many reactions contribute to the
ionization balance, and the major reactions vary over the transition layer. It is thus difficult to derive analytical formula
of electron abundances there. In the present work, we compare the electron abundance obtained by the equation (\ref{quad_e})
with the calculation of full reaction network at each position in the disk. We  adopt the latter,  when it is twice larger than the former.
As we have seen in the previous section, the position of the transition region depends sensitively
on the grain-size distribution in the disk model, as well as on the assumed UV flux from the central star and the
interstellar radiation fields. Readers are advised to use the PDR (photon-dominated region) codes to calculate the electron 
abundance in the AID layer
and transition region for a specific disk model. The PDR codes or their analogues are commonly used to calculate the vertical 
distributions of gas density and temperatures \citep[e.g.][]{kamp04, nomura05, gorti08}.

\subsection{H$_3^+$}
H$_3^+$ is produced by the reaction (\ref{h2p}).
The major destruction paths of H$_3^+$ are recombination and proton transfer to CO and N$_2$.
In the midplane at $R \lesssim$ a few 10 AU, the grain surface recombination also
becomes effective; 
\begin{eqnarray}
&& {\rm H_3^+ + e \rightarrow H_2 + H ~ ~ or~ ~ H + H + H} \label{h3p_rec} \\
&& {\rm H_3^+ + G(-) \rightarrow H_2 + H + G(0)~ ~ or~ ~ H + H + H + G(0)} \label{h3p_g} \\
&& {\rm{H}_3^+ + CO \rightarrow HCO^+ + H_2}, \label{hcop_form}\\
&& {\rm{H}_3^+ + N_2 \rightarrow N_2H^+ + H_2}, \label{n2hp_form} 
\end{eqnarray}

Considering the balance between the formation and destruction, H$_3^+$ abundance is given by
\begin{equation}
x({\rm H}_3^+)=\frac{1}{2}\frac{\zeta/n_{\rm H}}{k_{\ref{h3p_rec}}x({\rm e})+k_{\ref{h3p_g}}x({\rm G-})+k_{\ref{hcop_form}}x({\rm CO})+k_{\ref{n2hp_form}}x({\rm N}_2)}.
\label{H3p}
\end{equation}

\subsection{N$_2$H$^+$}
N$_2$H$^+$ is formed by the reaction (\ref{n2hp_form}), and is destroyed by the proton transfer to CO (reaction \ref{n2hp_co})
and recombination in the gas phase and on negatively charged grains:
\begin{eqnarray}
&& {\rm N_2H^+ + e \rightarrow  NH + N ~ ~ or ~ ~ N_2 + H} \label{n2hp_e}\\
&& {\rm N_2H^+ + G(-) \rightarrow  NH + N + G(0) ~ ~ or ~ ~ N_2 + H + G(0).} \label{n2hp_g} 
\end{eqnarray}
Then its abundance is given by
\begin{equation}
x({\rm N}_2{\rm H}^+)=\frac{k_{\ref{n2hp_form}} x({\rm H}_3^+)x({\rm N}_2)}{k_{\ref{n2hp_e}}x({\rm e})+k_{\ref{n2hp_co}}x({\rm CO})+k_{\ref{n2hp_g}}x({\rm G-})}. \label{n2hp}
\end{equation}

\subsection{HCO$^+$}
HCO$^+$ is formed by the reaction (\ref{n2hp_co}) and (\ref{hcop_form}),
and destroyed by the recombination in the gas phase (\ref{rec_hcop}) and on grain surfaces (\ref{hcop_g}).
Once the dust temperature exceeds $\sim 100$ K, molecules with higher proton affinity than CO, such as NH$_3$, are
desorbed to the gas phase to destroy HCO$^+$. In this section, we neglect such a high temperature
region, which is rather limited in our disk model. It is straightforward to derive
\begin{equation}
\frac{n(\rm{HCO}^+)}{n(\rm{CO})}=\frac{k_{\ref{hcop_form}}n({\rm H}_3^+)+k_{\ref{n2hp_co}}n({\rm N_2H^+})}{k_{\ref{rec_hcop}}n(\rm{e})+k_{\ref{hcop_g}}n({\rm G-})}.
\label{HCOp_analytic}
\end{equation}

At the disk surface region, where the grain surface recombination is not effective and H$_3^+$ dominates over N$_2$H$^+$,
the equation is modified to
\begin{equation}
\frac{n(\rm{HCO}^+)}{n(\rm{H}_3^+)}=\frac{k_{\ref{hcop_form}}n({\rm CO})}{k_{\ref{rec_hcop}}n(\rm{e})}.
\label{HCOp}
\end{equation}
It shows that HCO$^+$ is less abundant than H$_3^+$, if the abundance ratio of CO to electron is lower than
$\frac{k_{\ref{rec_hcop}}}{k_{\ref{hcop_form}}}=1.5\times 10^{2}(T/300 {\rm K})^{-0.69}$ (cf. \S 3.1). 



\subsection{Comparison with numerical results}
We now check how the analytical formulas compare with the full network calculation. We adopt the physical parameters
(density, temperature, and ionization rate) and abundances of CO and N$_2$ from the full network model at $t=1\times 10^5$ yr
and calculate
the abundances of electron, H$_3^+$, N$_2$H$^+$, and HCO$^+$ using the analytical formulas at each position
in the disk models with dark cloud dust and millimeter grains.
As described in \S 4.1., our analytical solution applies to a layer in which the major ions are molecular ions.
When the electron abundance obtained in the full network calculation is twice larger than given by the analytical formula
(i.e. AID layer),
we adopt the electron abundance from the former. It should be noted however, that the analytical formulas for
H$_3^+$, N$_2$H$^+$ and HCO$^+$ are appropriate even in the AID layer, if the electron abundance is adopted from the full network.

Figure \ref{analytic_2D} shows the 2D distributions of electron, H$_3^+$, N$_2$H$^+$ and HCO$^+$ abundances
calculated using the analytical formulas in the disk models with dark cloud dust (left panel) and millimeter grains (right panel).
The distributions of molecular ions are to be compared with those in the left columns of Figure \ref{ISM_chem} and
\ref{mm_chem}. 
The dashed lines indicate the height ($Z$) above which we adopt the electron abundance
of the full network calculation.
For a more quantitative comparison, Figure \ref{analytic_1D} shows the molecular distribution in the $Z$-direction at
$R=53.4$ AU in the disk models with dark cloud dust and millimeter grains. The solid lines
depict the abundances calculated by the full reaction network, while the dotted lines depict the analytical solution.

We can see that the analytical formulas are in reasonable agreement with the results of the full network calculation
in the both models.
At high $Z$ (e.g. $Z\gtrsim 0.4$ at $R\sim 30$ AU), H$_3^+$ is overestimated; while the analytical formula (\ref{H3p})
assumes that the hydrogen is all in H$_2$, it is photodissociated in the full network model
in such un-shielded low-density regions.
In molecular layers at lower $Z$, on the other hand, the analytic formulas tend to slightly overestimate
the molecular ion abundances, partly because the analytical formulas neglect the neutral species other than CO and N$_2$.
In the full network model, there are minor neutral species, such as OH,
which have a larger proton affinity than N$_2$ and CO.
In the upper layers of the millimeter grain model, the analytical formula underestimates
the HCO$^+$ abundance; another formation path, CO$^+$ + H$_2$ becomes effective in this region.


Figure \ref{analytic_col} shows the column densities of HCO$^+$ and N$_2$H$^+$ obtained by the full network calculation
in the millimeter grain model at $t=1\times 10^5$ yr (solid lines) and the analytical formulas (dotted lines).
The column densities of the analytical model agree with the full network model within a factor of 2. Although the analytical
formula underestimates the HCO$^+$ abundance at the disk surface, the surface region does not contribute much to the HCO$^+$
column density.

In summary, we have demonstrated that the analytical formulas of molecular ion abundances agree well with
the full network results. The analytical formulas give reasonable spatial distributions of molecular ions, taking into account
the spatial variation of gas density and ionization rate. They can thus be used to compare a disk model with molecular line
observations without performing the full chemical network calculations.
The input parameters of the formulas are density, temperature, ionization rate, and abundances of CO and N$_2$.
The distributions of density and temperature should be prepared for a specific object.
Then the ionization rate can be calculated by X-ray radiation transfer and by assuming a cosmic-ray penetration depth and abundances
of radioactive nuclei \citep[e.g.][]{cleeves14}. The electron abundance in the disk surface is also needed, and can be obtained
by the PDR calculation. Alternative option is to simply assume a height from the midplane below which molecular ions are more
abundant than atomic ions.
Finally we need spatial distributions of CO and N$_2$. While the full network calculations show that they are subject to
the sink effect and thus could decrease with time even in the region warmer than their sublimation temperature, the simplest
assumption would be the equilibrium abundance between adsorption and desorption, as we will see in the next section.

\section{Results: No Sink Model}

In our full network calculations, N$_2$H$^+$ ring is reproduced in a millimeter grain model with a significant CO depletion due
to the sink effect. Efficiency of the sink effect, however, depends on various parameters.
It is therefore useful to calculate the distribution of N$_2$H$^+$ in a model without the sink effect.
As described in \S 2.3, here we assume that the total (gas and ice) abundances of CO and N$_2$ are constant,
and that their gas/ice ratios
are determined by the equilibrium between the adsorption onto and desorption from the grain surfaces
(eq. \ref{2phase} and \ref{CO_analytical}).

A combination of the analytical formulas of molecular ions and equilibrium abundances of CO and N$_2$ make it very easy to investigate
the dependence of N$_2$H$^+$ abundance on various disk parameters. Here we demonstrate this merit of the analytical formulas by
investigating the dependence of N$_2$H$^+$ abundance on desorption energies of CO and N$_2$, and on ionization rate in the disk
model with millimeter grains.
We could apply the no-sink model to the disk model with dark cloud dust, as well, but the relatively high abundance of N$_2$H$^+$
in the disk surface makes the radial distribution of N$_2$H$^+$ column density flatter than observed (\S 3.1).

\subsection{N$_2$H$^+$ in Disk Models Without Sink Effect}

Figure \ref{analytic_2D_CON2} ($a-c$) shows the distributions of CO, N$_2$, and N$_2$H$^+$ in the no-sink model
in the disk model with millimeter grains. The desorption energies of CO and N$_2$ are assumed to be 1150 K and 1000 K,
respectively. The dashed line depicts the height above which the electron abundance is adopted from the full-network model. 
Figure \ref{analytic_2D_CON2} ($d-e$) shows the vertical distributions of molecules at the radius of 40 AU and 93 AU.
At $R=40$ AU, the temperature is close to the sublimation temperatures of CO and N$_2$ even in the midplane. The gas-phase
abundances of CO and N$_2$ are thus mostly determined by the thermal desorption. In the midplane at $R=93$ AU, on the other hand,
the non-thermal desorption by cosmic-ray dominates over the thermal desorption. 
The abundance of N$_2$H$^+$ is determined by the equation (\ref{n2hp}). As expected, N$_2$H$^+$ abundance is high in layers
where N$_2$ is more abundant than CO. We note however, that N$_2$H$^+$ abundance is determined not only by the abundance
ratio of N$_2$/CO. For example, in Figure \ref{analytic_2D_CON2} ($d-e$), the N$_2$H$^+$ abundance varies 
among the three positions where the abundances of CO and N$_2$ are equal. The absolute values of CO and N$_2$ abundances matter, since
they control the abundance of H$_3^+$ (eq. \ref{H3p}), from which N$_2$H$^+$ is formed. 
Note that N$_2$H$^+$ abundance in the very surface region of the disk (e.g. $Z\gtrsim 0.3$ at $R\sim 50$ AU) could be overestimated;
its mother molecule, N$_2$, is photodissociated in the disk surface in the full network model.

The radial distributions of N$_2$H$^+$ column density is shown in Figure \ref{analytic_2D_CON2} ($f$).
The dotted line depicts the CO column density multiplied by a factor of $10^{-7}$. 
In order to avoid the photodissociation region at the disk surface, N$_2$H$^+$ column density is calculated by the integration
at $|Z|\le 0.3$ ($R > 50$ AU), $|Z|\le 0.15$ ($10 < R \le 50$ AU), and $|Z|\le 0.1$ ($R\le 10$ AU), while the CO column density
is obtained by the integration along the whole disk height. We can see that N$_2$H$^+$ column density indicates a ring structure,
as in the full-network model. The spatial distribution of N$_2$H$^+$ is however, different
from that in the full-network model. 
Firstly, the N$_2$H$^+$ ring is sharp and exists right outside the CO snow line in the no-sink
model, simply reflecting the lower desorption energy of N$_2$ than that of CO, while it is broader in the full-network model, especially in the late stage ($t=9.3\times 10^5$ yr).
The peak N$_2$H$^+$ column density at $R \sim 50$ AU
and $t > 10^5$ yr is lower in the full network model, because N$_2$ is depleted via the sink effect in the midplane. 
Secondly, N$_2$H$^+$ is confined to a thin layer colder than the CO sublimation temperature
in the no-sink model, while it is abundant also in the upper warmer layers in the full-network model.
The two models can thus be distinguished by constraining the vertical
distributions of N$_2$H$^+$ from observations. 

In the case of TW Hya, N$_2$H$^+$ lines of $J=3-2$  and $J=4-3$ are observed \citep{qi13sma,qi13}. Assuming
that the lines are optically thin under LTE conditions, the excitation temperature is derived to be $35\pm10$ K. The relatively high excitation temperature
of N$_2$H$^+$, together with the low CO abundance indicated by C$^{18}$O and HD observations \citep{favre13}, might be
better explained by the full-network model. We postpone the discussion on TW Hya to a future work, in which
we will apply our network model and analytical formulas to a disk model specified for TW Hya, and simulate molecular emission with non-LTE radiative
transfer calculations for more quantitative comparisons with the observational data.

\subsection{Dependence on desorption rates}
So far, we set the desorption energies of CO and N$_2$ to be 1150 K and 1000 K, respectively.
It is well known however, that the desorption energies of molecules depend on the chemical compositions
and physical structure (e.g. crystal or amorphous) of the ice mantle. While the values we adopted 
from \cite{garrod06} are desorption energies on water-dominated ice surfaces derived from the temperature programmed desorption (TPD)
experiments \citep{collings04}, the desorption energies of pure ices
of CO and N$_2$ are $E_{\rm des}$(CO)$=855\pm 25$ K and $E_{\rm des}$(N$_2)=790\pm 25$ K \citep{oberg05}. 
In \S 3.2, we have shown that in the full-network model the N$_2$H$^+$ distribution does not sensitively depend on
the desorption energies of CO and N$_2$, because the layer of CO depletion, where N$_2$H$^+$ is abundant, is determined
by the sink effect rather than thermal desorption.
If the CO sink is not effective, on the other hand, the abundance of N$_2$H$^+$ could be more sensitive to their
desorption energies.

We also note that the equation (\ref{2phase}) assumes the first-order desorption; i.e. the desorption rate is proportional to the 
abundance of the species in the ice mantle, $n_{\rm CO ice}$. In the following, we call this model as "2 phase", since this
equation is usually used in the 2-phase gas-grain chemical models, which do not discriminate the layers of ice mantle.
In reality, the migration (and thus desorption) of molecules deeply embedded in the ice mantle could be inefficient,
at least at low temperatures \citep[e.g.][]{collings04}.
We therefore consider another model, i.e. "3-phase" model, in which only the surface monolayer is subject to desorption:
\begin{equation}
S \pi a^2  v_{\rm th}n_{\rm COgas} n_{\rm{dust}}=\min[4\pi a^2 n_{\rm{dust}} N_{\rm site}, n_{\rm CO ice}]
\Bigl\{\nu \exp \Bigl(-\frac{E_{\rm des}^{\rm CO}}{kT}\Bigr)+\nu\tau_{\rm CR}C_{\rm Fe}
\exp \Bigl(-\frac{E_{\rm des}^{\rm CO}}{kT_{\rm max}}\Bigr)\Bigr\}, \label{eq:3phase}
\end{equation}
where $N_{\rm site}=1.5\times 10^{15}$ cm$^{-2}$ is the number density of adsorbing site on a grain surface. The desorption is
zero-th order, as long as the ice is abundant enough to occupy more than a monolayer in the ice mantle,
i.e. $n_{\rm CO ice} > 4\pi a^2 n_{\rm{dust}} N_{\rm site}$. Then the gaseous abundance is independent of the 
total abundance. In the 2-phase model, on the other hand, the gaseous CO (N$_2$) abundance is proportional to the assumed total 
abundance of CO (N$_2$).

Before showing the N$_2$H$^+$ abundance in the disk model with various desorption rates, it is instructive to apply
the analytical formulas to a simpler model, where the gas density and ionization rate are constant, $10^8$ cm$^{-3}$ and $5\times
10^{-17}$ s${^-1}$, respectively. Figure \ref{N2Hp_T} shows the gaseous abundances of CO, N$_2$ and N$_2$H$^+$ as a function of
temperature. The gas and dust temperatures are set to be equal.
The total (gas and ice) abundances of CO and N$_2$ are $1\times 10^{-4}$ and $4.5\times 10^{-6}$, respectively.
The desorption energies are set to be $E_{\rm des}$(CO)$=E_{\rm des}$ (N$_2$)$=855$ K in Figure \ref{N2Hp_T}(a),
and $E_{\rm des}$(CO)$=855$ K and $E_{\rm des}$(N$_2$)$=790$ K in Figure \ref{N2Hp_T}(b). The desorption rate of 2-phase
model (eq. \ref{2phase}) is assume for the red lines, while the 3-phase model is assume for the blue lines.
The abundances change drastically at the sublimation temperature $\sim 20$ K.
At $T\lesssim 15$ K, the gaseous abundances of CO and N$_2$ slightly increase with decreasing temperature;
desorption rate is kept constant due to the non-thermal desorption,
while the adsorption rate is proportional to $T^{1/2}$.
By comparing the red and blue lines,
we can see that the sublimation temperature and gaseous abundances of CO and N$_2$ at low temperatures are
significantly different between the 2-phase and 3-phase models; the 3-phase model gives much smaller desorption rate than
the 2-phase model, when the ice mantle is thick.
In the 3-phase model with the same $E_{\rm des}$ for CO and N$_2$ (blue lines in panel a), the gaseous abundance
of CO and N$_2$ are the same below the sublimation temperature. It is interesting that the N$_2$H$^+$ abundance shows
a sharp peak slightly below the sublimation temperature.
It clearly shows that the N$_2$H$^+$ abundance
depends not only on the abundance ratio of CO/N$_2$, but also on the absolute values of their abundances.
Specifically, the N$_2$H$^+$ abundance reaches the maximum value when the abundance ratio of CO to electron is
$\sim k_{\ref{n2hp_e}}/k_{\ref{n2hp_co}}$, which is about $3.3\times 10^3$ at $T=17$ K (see Appendix B).
In the models of Figure \ref{N2Hp_T}, the electron abundance is about several $\times 10^{-10}$ and varies slightly with
temperature and desorption model. At warm temperatures where the abundance ratio of CO to electron is higher than this
critical value, HCO$^+$ is the dominant ion,
while H$_3^+$ dominates at lower temperatures.
In the case of 2-phase model with the same $E_{\rm des}$ for CO and N$_2$(red lines), on the other hand,
N$_2$H$^+$ abundance does not show such a sharp peak around the CO sublimation temperature,
because the abundance ratio of CO to electron is always higher than the critical value mentioned above.
In the case of $E_{\rm des}$ (N$_2$) $<$  $E_{\rm des}$ (CO) (Figure \ref{N2Hp_T} b),
there is a narrow temperature range in which N$_2$ is relatively abundant but CO is not, so that N$_2$H$^+$ abundance
has the maximum value, even in the 2-phase model.

Figure \ref{analytic_col_CON2} shows the radial distributions of N$_2$H$^+$ (solid lines) and CO (dotted lines) column densities
in the disk model with millimeter grains as in Figure 9 ($f$), but for three sets
of desorption energies of CO and N$_2$: $E_{\rm des}$(CO)=1150 K and $E_{\rm des}$(CO)=1000 K in panel (a), 
$E_{\rm des}$(CO)=855 K and $E_{\rm des}$(CO)=790 K in panel (b), and $E_{\rm des}$(CO)=855 K and $E_{\rm des}$(CO)=855 K
in panel (c). The column density of hydrogen nuclei is multiplied by a factor of $10^{-11}$, and is shown
by the green line. The black and red lines depict the 2-phase and 3-phase models, respectively.
The total (gas and ice) abundance of N$_2$ is $4.5\times 10^{-6}$ relative to hydrogen for the solid
lines. In order to investigate the dependence of N$_2$H$^+$ abundance on the total N$_2$ abundance, it is set to be
10 times higher for the dashed lines. We can see that all the models show a ring-like
structure of N$_2$H$^+$. The N$_2$H$^+$ column density and its radial gradient however, depend significantly on the desorption
energies and the model of desorption. The CO snow line and the inner edge of N$_2$H$^+$ ring are at smaller radii in the model
with higher $E_{\rm des}$(CO). They are also at smaller radii in the 3-phase model than in the 2-phase model, because of the lower
desorption rate in the 3-phase model.
While the N$_2$H$^+$ column density significantly depends on the total N$_2$
abundance in the 2-phase model, the dependence is weak in the 3-phase model.
Dependence of the N$_2$H$^+$ column density on the desorption energies of CO and N$_2$ are complex, as expected from
Figure \ref{N2Hp_T}.
In the models with $E_{\rm des}$(CO) $> E_{\rm des}$ (N$_2$), there is a region where N$_2$ is thermally desorbed but CO is
not, which results in a peak of N$_2$H$^+$ column density with a width of a few 10 AU. 
In the models with $E_{\rm des}$(CO) $ = E_{\rm des}$ (N$_2$)$=855$ K, the N$_2$H$^+$ column density shows a sharp peak in the 3-phase
model, but not in the 2-phase model, as expected from Figure \ref{N2Hp_T}.


\subsection{Dependence on Ionization rate}
Although we have assumed that the attenuation length of the cosmic ray ionization is 96 g cm$^{-2}$\citep{umebayashi81},
the stellar winds and/or magnetic fields could hamper the penetration of cosmic rays to the disk.
Then X-ray would be the major ionization
source \citep{cleeves13, glassgold97}. The dotted lines in Figure \ref{mm_chem} depict the height from the midplane
below which the X-ray ionization rate is lower than $10^{-17}$ s$^{-1}$ and $10^{-18}$ s$^{-1}$, respectively.
The layer with abundant N$_2$H$^+$ are around or below these lines in both the full network model (i.e. with CO sink)
and the no-sink model.
Here we investigate how the N$_2$H$^+$ column density changes, if the cosmic ray does not reach the disk.

Figure \ref{fig12} shows the radial distribution of N$_2$H$^+$ column density without the cosmic-ray ionization.
X-ray is the only ionization source for the dashed lines, while the decay of radioactive nuclei is considered for the solid lines.
It should be noted that cosmic ray also causes non-thermal desorption in our models; the non-thermal desorption by the cosmic-ray
is neglected for the solid lines and dashed lines.
For a comparison, the dotted lines depict the model in which the non-thermal desorption is included but
ionization source is X-ray and radioactive nuclei; although such a model is not self-consistent, we can see the importance
of the non-thermal desorption by comparing the solid lines with dotted lines.
The blue and green lines depict the model in which CO and N$_2$
abundances in the gas phase are given by the 3-phase model; labels in the figure depict the assumed desorption energies of CO and N$_2$.
For the red lines, we refer to the CO and N$_2$ abundances in our full network model at $1\times 10^5$ yr, i.e. the model with sink.
Note that there is thus no red dotted line.
Compared with the models with cosmic ray ionization (Figure \ref{analytic_col} and \ref{analytic_col_CON2}), the N$_2$H$^+$ column density is
reduced significantly. The reduction factor, the ratio of the peak column density in the model with cosmic ray to that in the
model without cosmic-ray, is the highest ($\sim 70$) in the no-sink model with low CO and N$_2$ desorption energies,
while it is the lowest (factor of $\sim 20$) for the model with sink. In the latter model, N$_2$H$^+$ is relatively abundant
even in the layer with the X-ray ionization rate $\gtrsim 10^{-18}$ s$^{-1}$ (i.e. $Z/R \gtrsim 0.2$). In the no-sink models,
on the other hand, N$_2$H$^+$ is depleted at $0.2\lesssim Z\lesssim 0.3$. 
In the model with lower desorption energy of CO, the CO freeze-out region, where the N$_2$H$^+$ is abundant, is confined
to the layer closer to the midplane with low X-ray ionization rate.

\section{Summary}

In this work we calculated the molecular abundances in disk models
to investigate the origin of N$_2$H$^+$ ring recently found at the CO snow line in the disk of TW Hya \citep{qi13}.
We adopt two disk models with different dust properties; dust grains are similar to dark cloud dust in one model, while
they have grown to millimeter size in the other model. We first calculated a full network
of gas-grain chemical reactions. 
Our findings in the full-network model are as follows.
\begin{itemize}
\item{In the model with dark cloud dust, N$_2$H$^+$ is abundant in the disk surface.
Although N$_2$H$^+$ abundance near the midplane has a local maximum outside the radius of CO snow line,
N$_2$H$^+$ column density is rather high even inside the CO snow line, because N$_2$H$^+$ in the disk surface
contributes significantly to the column density.}
\item{In the model with millimeter grains, the column density of N$_2$H$^+$ shows a local peak around the CO
snow line. N$_2$H$^+$ is abundant in the warm intermediate layer where CO is depleted via the sink effect,
while N$_2$H$^+$ is destroyed by the proton transfer to CO inside the CO snow line.
Penetration of UV radiation (i.e. lower extinction than the dark cloud dust model) plays two important roles.
Firstly, UV radiation makes the atomic ions dominant in the disk surface, so that N$_2$H$^+$ is not abundant
there. In the intermediate layer where CO is depleted via the sink effect, the conversion of N$_2$ to NH$_3$ (i.e. the sink effect on N$_2$)
is prevented by the photodissociation of NH, so that N$_2$H$^+$ is abundant.}
\item{In the model with millimeter grains, the distributions of N$_2$H$^+$ and its column density do not change significantly
when the desorption energies of CO and N$_2$ are varied. The N$_2$H$^+$ column density shows a local peak at the radius of CO snow line,
even if the desorption energies of CO and N$_2$ are equal, because in the region with abundant N$_2$H$^+$, CO is depleted via the
sink effect rather than the adsorption of CO itself onto grain surfaces.}
\item{By analyzing the results of the full-network model, we derived analytical formulas of electron, H$_3^+$, N$_2$H$^+$ and HCO$^+$
abundances as functions of gas density, temperature, ionization rate, and abundances of CO and N$_2$.
The analytical formulas would be useful for radio observers
to derive the abundances of these molecular ions from the observational data using a reasonable physical model
for a well-observed disk such as TW Hya.}
\end{itemize}

While the sink effect on CO plays an important role in determining the N$_2$H$^+$ abundance in the full-network model,
the efficiency of the sink depends on various parameters such as turbulence in the disk and the rates
of chemical conversion of CO to less volatile species. We thus constructed the no-sink model, in which the total
(gas and ice) abundances of CO and N$_2$ are set constant, and their gas-phase abundances are determined by the balance between
the adsorption and desorption. The abundances of molecular ions are calculated using the analytical formulas.
The results of the no-sink models are as follows.

\begin{itemize}
\item{The column density of N$_2$H$^+$ shows a ring-like structure in the no-sink model with millimeter dust grains.
Since the abundance of N$_2$H$^+$ is given by a non-linear function
of CO and N$_2$ abundances, it can reach a moderate value even in the cold N$_2$ freeze-out region, depending on the
abundances of CO and N$_2$.}
\item{ Even if the desorption energies of CO and N$_2$ are the same, N$_2$H$^+$ abundance peaks at the temperature
slightly below the CO (and N$_2$) sublimation temperature, where the abundance ratio of CO to electron is $\sim 
k_{\ref{n2hp_e}}/k_{\ref{n2hp_co}}$ $\sim 10^3$.}
\item{Although the N$_2$H$^+$ ring is produced both in the full network model and in the no-sink model,
the detailed distributions of N$_2$H$^+$ are different between the two models. In the no-sink model,
N$_2$H$^+$ abundant layer is confined to a layer colder than
the sublimation temperature of CO, while in the full-network model (i.e. with CO sink), N$_2$H$^+$ abundant layer extends
to warmer layers. 
These models can thus be discriminated in observations if we can determine the vertical distribution of N$_2$H$^+$ in disks, specifically,
starting with constraining the excitation temperature of N$_2$H$^+$ emission. }
\item{The column density of N$_2$H$^+$ in the no-sink model sensitively depends on the desorption rates of
CO and N$_2$. } 
\item{If the cosmic ray does not reach the disk, the N$_2$H$^+$ column density is reduced by a factor of 20 in the
model with CO sink, and by a factor of 70 in the no-sink model. The reduction is more significant in the model with lower
desorption energy of CO, because N$_2$H$^+$ (i.e. CO freeze-out) is confined to a layer closer to the midplane,
where X-ray ionization rate is lower.}
\end{itemize}




\acknowledgments

We thank Karin Oberg and Catherine Walsh for helpful discussions. We thank the anonymous referee for his/her
constructive comments.
This work was supported by JSPS KAKENHI Grant Numbers 23103004, 23103005, 23540266 and 25400229.
Numerical calculation is partly conducted using SR16000 at YITP in Kyoto University for numerical calculations.
K.F. is supported by the Research Fellowship from the Japan Society for the Promotion of Science (JSPS).
\appendix

\section{Reaction Rate Coefficients}
We list the rate coefficients of the reactions relevant to the analytical formulas of molecular ions
in Table \ref{tab_reactions}. In the full network model, the spin state (ortho and para) of H$_3^+$
is discriminated, and the reaction rates actually vary with the spin state. For those reactions,
we use the rate coefficients from \cite{garrod06} in the analytical formulas, for simplicity.

\begin{table}
\begin{center}
\caption{Reaction rate coefficients relevant to the analytical formulas\tablenotemark{a}}
\begin{tabular}{lccc}
\tableline\tableline
reaction & $\alpha$ & $\beta$\\ 
\tableline
${\rm H_3^+ + CO \rightarrow HCO^+ + H_2}$ & 1.61(-9)\tablenotemark{b} & 0.0  \\
${\rm H_3^+ + N_2 \rightarrow  N_2H^+ + H_2}$ & 1.7(-9) & 0.0 \\
${\rm H_3^+ + e \rightarrow H_2 + H}$ & 2.59(-8) & -0.5 \\
${\rm H_3^+ + e \rightarrow H + H + H}$ & 4.61(-8) & -0.5  \\
${\rm HCO^+ + e \rightarrow H + CO}$ & 2.40(-7) & -0.69 \\
${\rm N_2H^+ + CO \rightarrow  HCO^+ + N_2}$ & 8.8(-10) & 0.0 \\
${\rm N_2H^+ + e \rightarrow NH + H}$ &  1.30(-8) &  -0.84  \\
${\rm N_2H^+ + e \rightarrow  N_2 + H}$ &  2.47(-7) &  -0.84 \\
\tableline
\end{tabular}
\label{tab_reactions}
\tablenotetext{a}{Rate coefficients are given in the format $k=\alpha \times (T/300.0)^{\beta}$ cm$^3$ s$^{-1}$.}
\tablenotetext{b}{$A(B)$ stands for $A\times 10^{B}$}
\end{center}
\end{table}

\section{Dependence of N$_2$H$^+$ Abundance on Temperature}

Figure \ref{N2Hp_T} shows that N$_2$H$^+$ abundance reaches the maximum value at $T\sim 17$ K in the 3-phase
model, even if we assume the same desorption energy for CO and N$_2$. Here we analyze the model to show that N$_2$H$^+$
abundance is maximized when the abundance ratio of CO to electron is $\sim 10^3$.

At the density of $n_{\rm H}=10^8$ cm$^{-3}$,
gas-phase recombination is more effective than the grain surface recombination. Then the analytic formula of the N$_2$H$^+$
abundance (eq. \ref{n2hp}) becomes
\begin{equation}
x({\rm N}_2{\rm H}^+)=\frac{k_{\ref{n2hp_form}} x({\rm H}_3^+)x({\rm N}_2)}{k_{\ref{n2hp_e}}x({\rm e})+k_{\ref{n2hp_co}}x({\rm CO})}. \label{n2hp_app}
\end{equation}
When the abundance ratio of CO to electron is higher than $k_{\ref{n2hp_e}}/k_{\ref{n2hp_co}}$, which is about $3.3\times 10^3$
at $T=17$ K, the denominator is dominated by the second term. Substituting the analytical formula of H$_3^+$ abundance
(eq. \ref{H3p}) to (\ref{n2hp_app}), it is straightforward to show that
N$_2$H$^+$ abundance is proportional to $x$(N$_2$)/$x^2$(CO), and thus increases with decreasing CO
abundance at $T\sim 17-20$ K.
If the first term dominates in the denominator, on the other hand, N$_2$H$^+$ abundance
is proportional to N$_2$ abundance. The N$_2$H$^+$ abundance decreases with decreasing temperature at $T\sim 15-17$ K.

\clearpage



\begin{table}
\begin{center}
\caption{Elemental abundance and initial abundances of assorted molecules}
\begin{tabular}{llll}
\tableline\tableline
Element & Abundance\tablenotemark{a} & Element & Abundance\\ 
\tableline
H & 1.0 & He & 9.75(-2) \\
N & 2.47(-5) & O & 1.80(-4) \\
C & 7.86(-5) & S & 9.14(-8) \\
Si & 9.74(-9) & Fe & 2.74(-9) \\
Fe & 2.74(-9) & Na & 2.25(-9) \\
Mg & 1.09(-8) & & \\
\tableline
Species & Abundance & Species & Abundance \\
H$_2$O & 1.15(-4) & CO & 3.57(-5) \\
CO$_2$ & 3.52(-6) & CH$_4$ & 1.50(-5) \\
H$_2$CO & 1.20(-5) & CH$_3$OH & 6.50(-6) \\
N$_2$ & 4.47(-6) & NH$_3$ & 1.44(-5) \\
\tableline
\end{tabular}
\label{tab_init_abun}
\tablenotetext{a}{$A(-B)$ means $A \times 10^{-B}$.}
\end{center}
\end{table}

\begin{figure}
\epsscale{.80}
\plotone{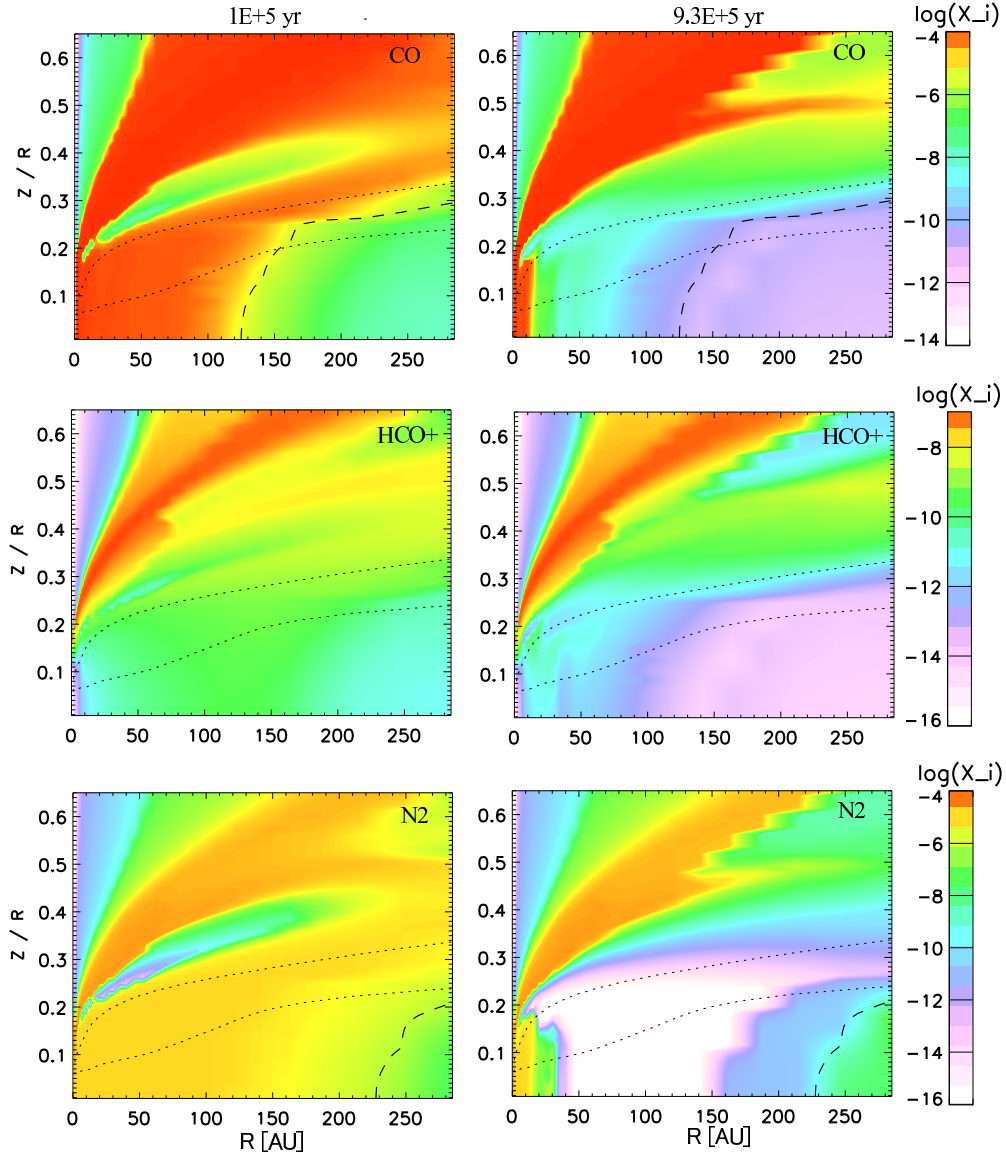}
\caption{Distributions of CO, HCO$^+$, N$_2$, N$_2$H$^+$, H$_3^+$ and electron in the gas phase at $t=1\times 10^5$ yr
(left panels) and $9.3\times 10^5$ yr (right panels) in the model with dark cloud dust.
Dashed lines in the panels of CO and N$_2$ depict the position where the gas-phase and ice-mantle abundances become
equal in the adsorption-desorption equilibrium (eq. \ref{snowline}).
The dotted lines depict the position where the X-ray ionization rate
is equal to the cosmic-ray ionization rate ($5\times 10^{-17}$ s$^{-1}$) and to the ionization rate by decay of radioactive nuclei
($1\times 10^{-18}$ s$^{-1}$).
\label{ISM_chem}}
\end{figure}

\setcounter{figure}{1}
\begin{figure}
\plotone{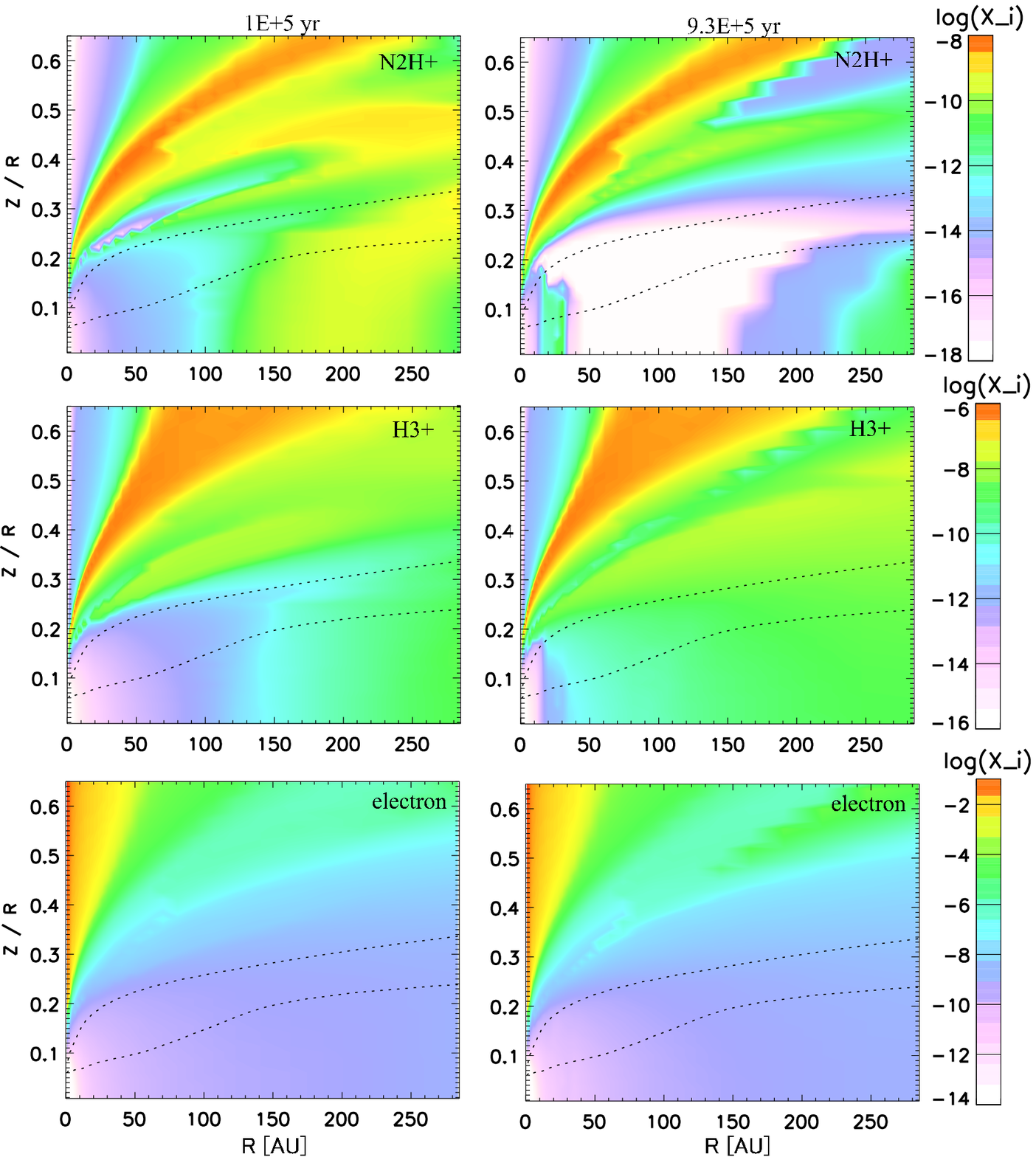}
\caption{cont. 
}
\end{figure}

\begin{figure}
\epsscale{.80}
\plotone{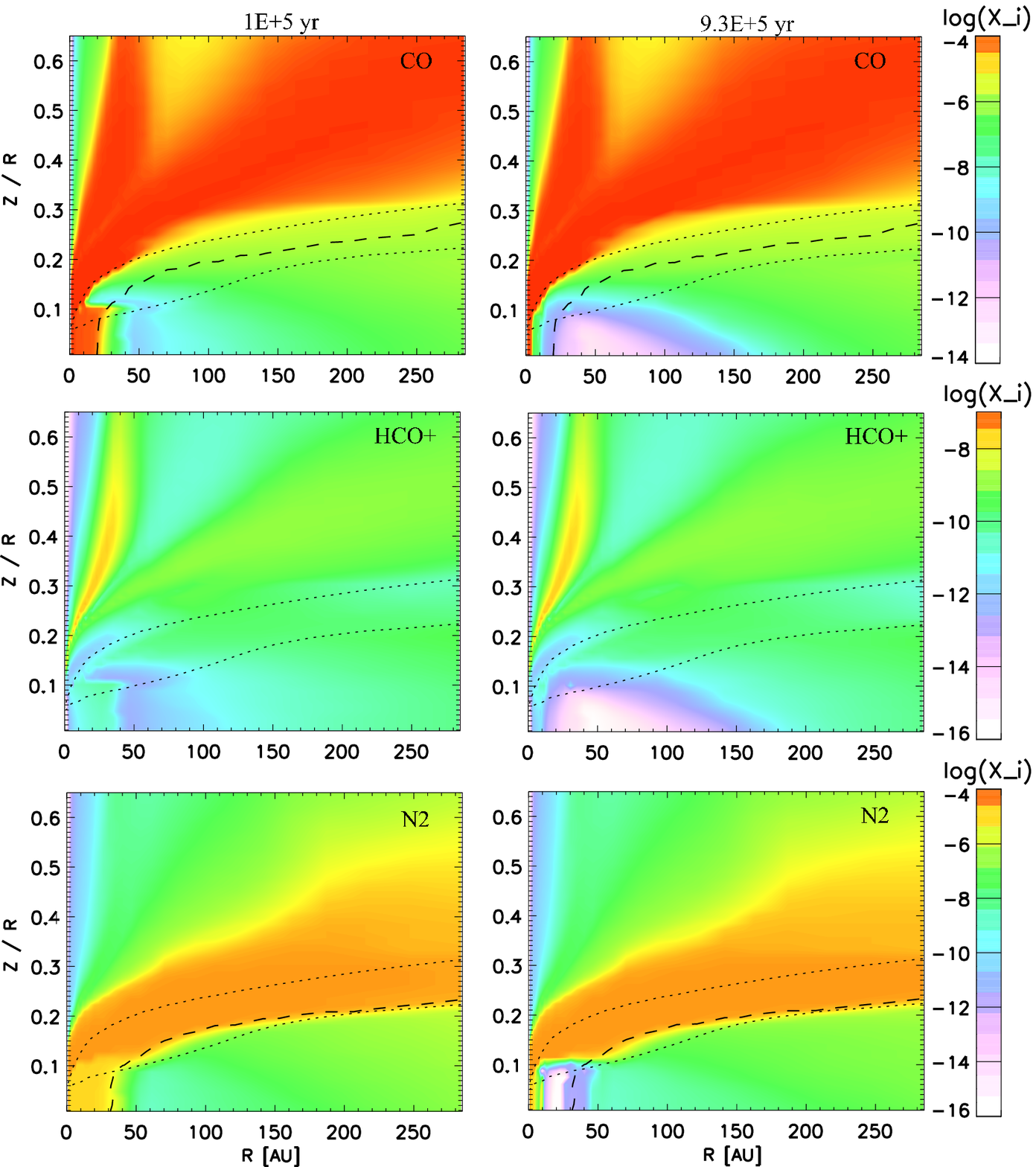}
\caption{Distributions of CO, HCO$^+$, N$_2$, N$_2$H$^+$, H$_3^+$ and electron
as in Figure 2, but for the model with millimeter grains.
\label{mm_chem}}
\end{figure}

\setcounter{figure}{2}
\begin{figure}
\epsscale{0.8}
\plotone{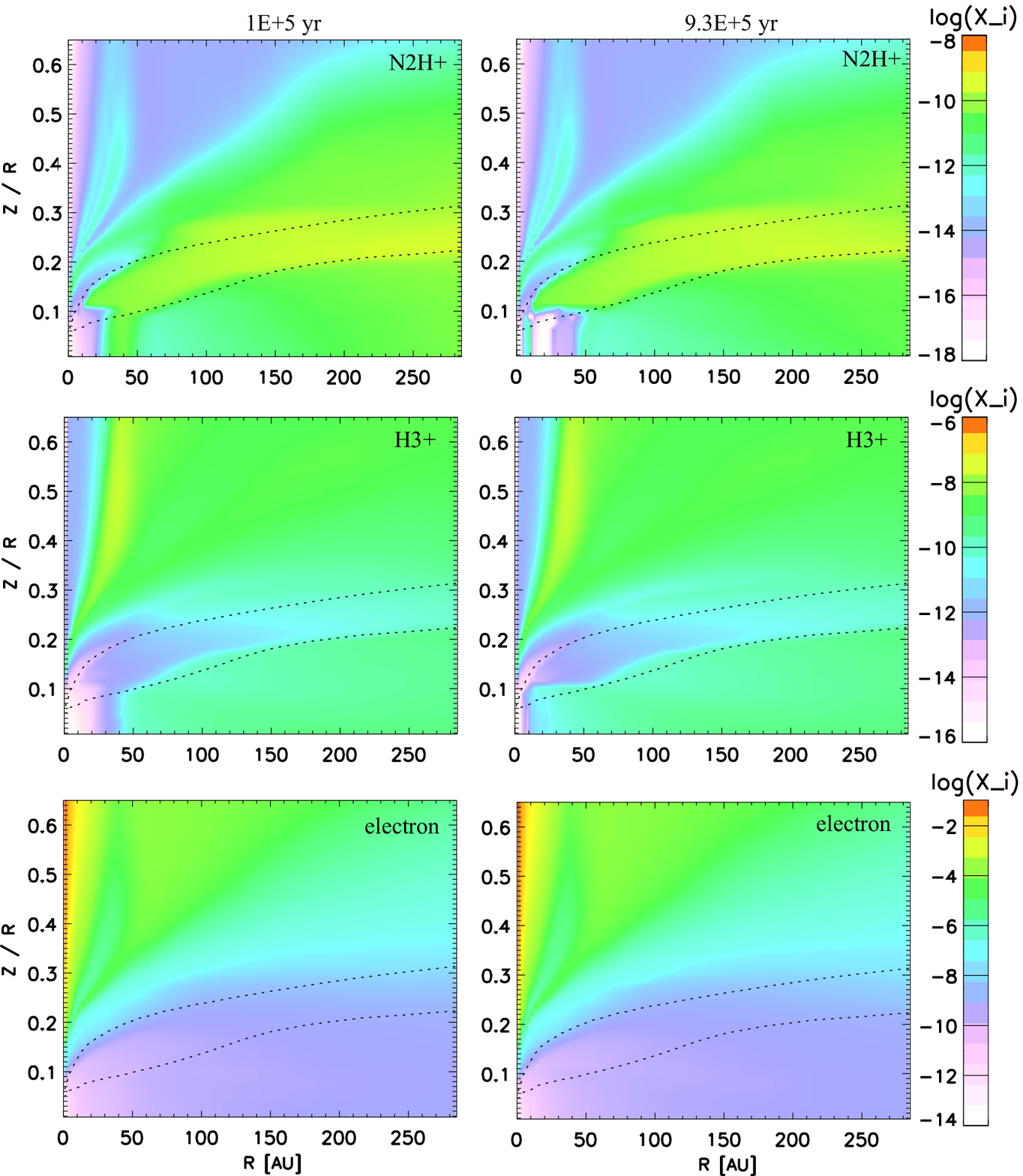}
\caption{cont.
}
\end{figure}

\begin{figure}
\epsscale{.80}
\plotone{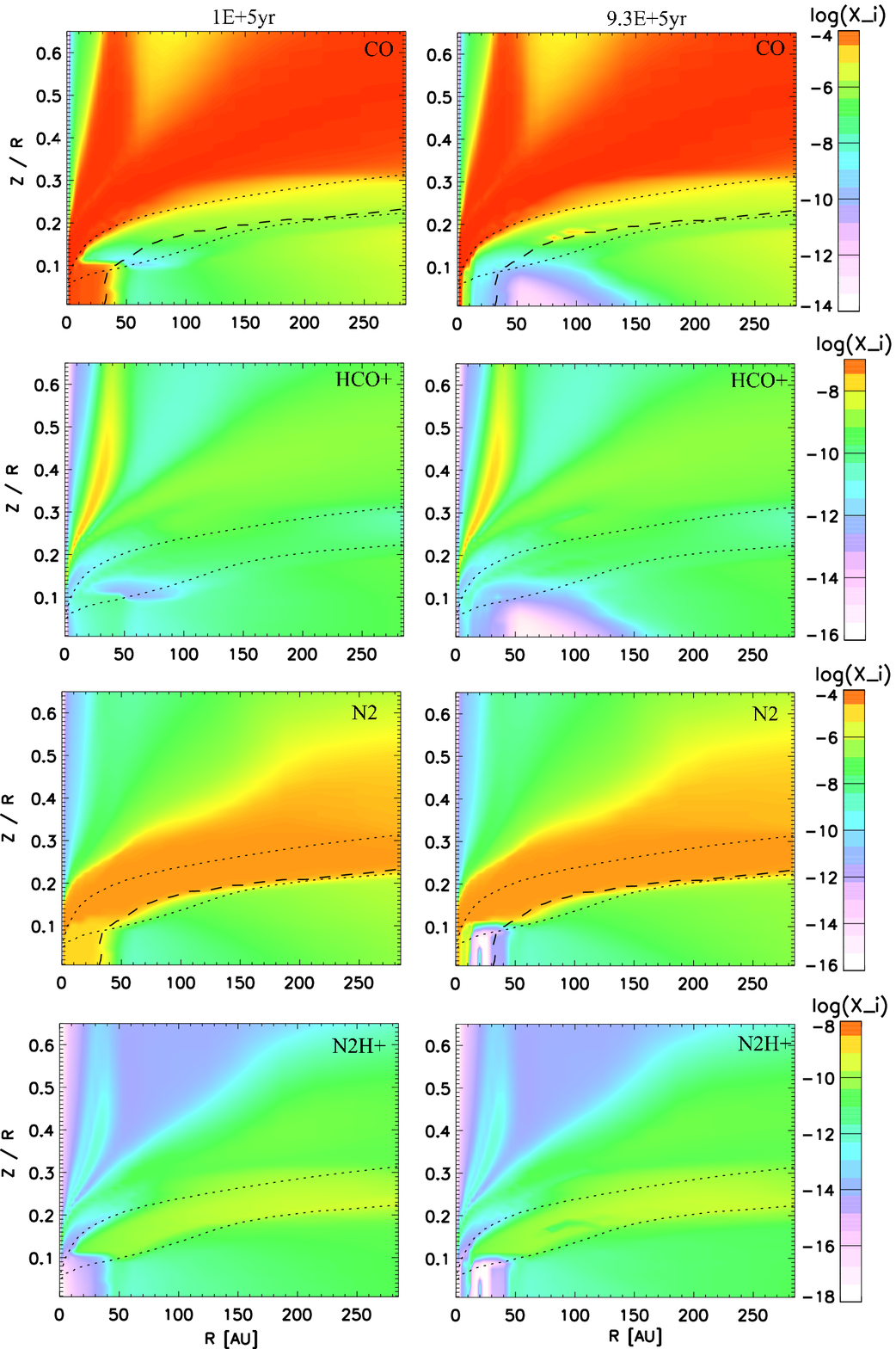}
\caption{Distributions of CO, HCO$^+$, N$_2$, and N$_2$H$^+$ in the model with millimeter grains
as in Figure \ref{mm_chem}. The desorption energies of CO and N$_2$ are set to be equal (1000 K).
\label{CO1000_chem}}
\end{figure}

\begin{figure}
\plotone{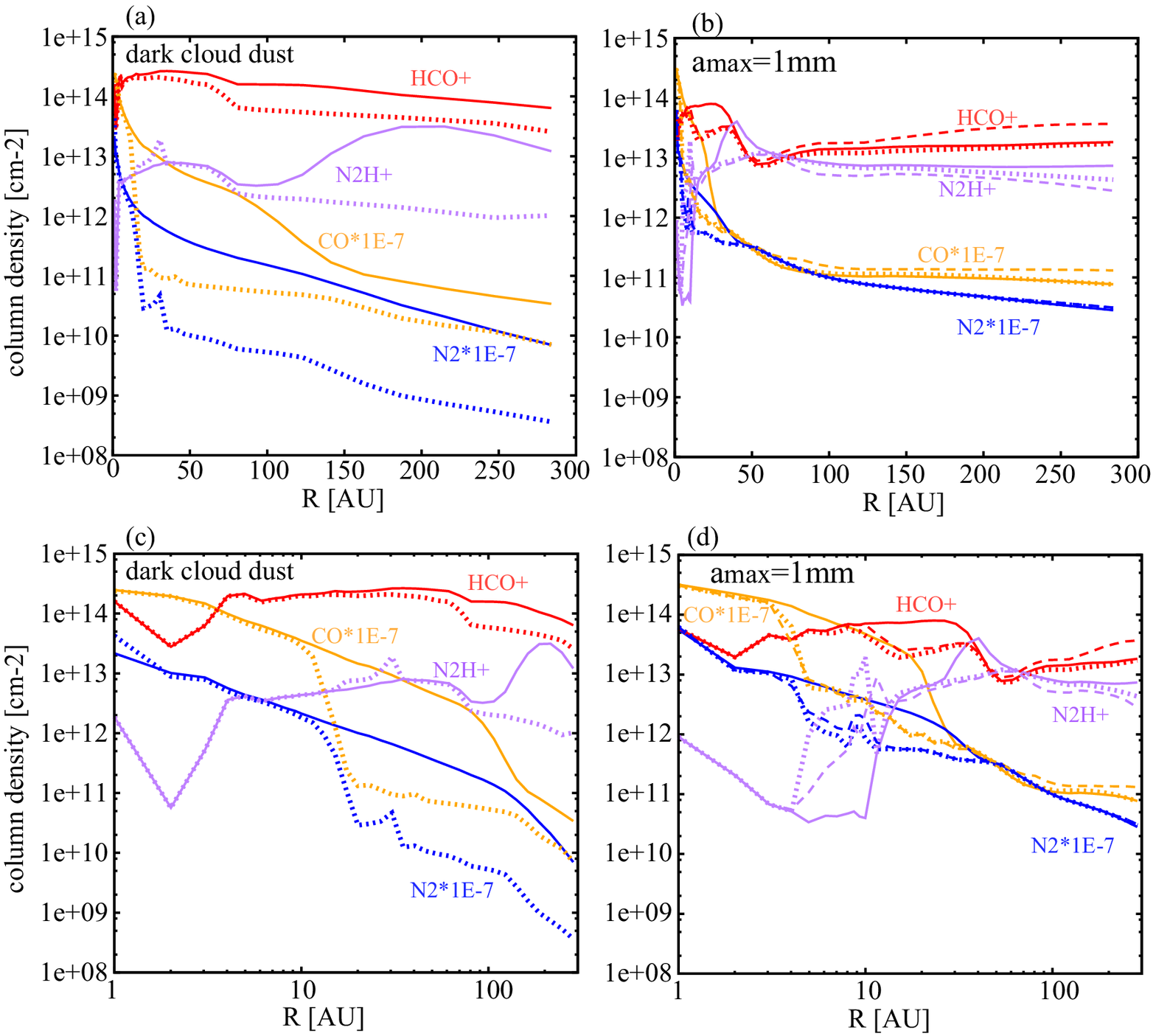}
\caption{Radial distributions of column densities of CO, HCO$^+$, N$_2$ and N$_2$H$^+$ in the gas phase
at $t=1\times 10^5$ yr (solid lines) and $9.3\times 10^5$ yr (dotted lines) in the model with dark cloud dust
(a and c) and millimeter grains (b and d). The column densities of CO and N$_2$ are multiplied by a factor of
$10^{-7}$ to fit in the figure.
The horizontal axis, $R$, is linear in (a) and (b),
while it is logarithmic to highlight the inner radius in (c) and (d).
The dashed lines in the right panel depict the column densities
of molecules in the model with $E_{\rm des}$(CO)$=E_{\rm des}$(N$_2$)=1000 K at $9.3\times 10^5$ yr.
\label{column}}
\end{figure}

\begin{figure}
\epsscale{.80}
\plotone{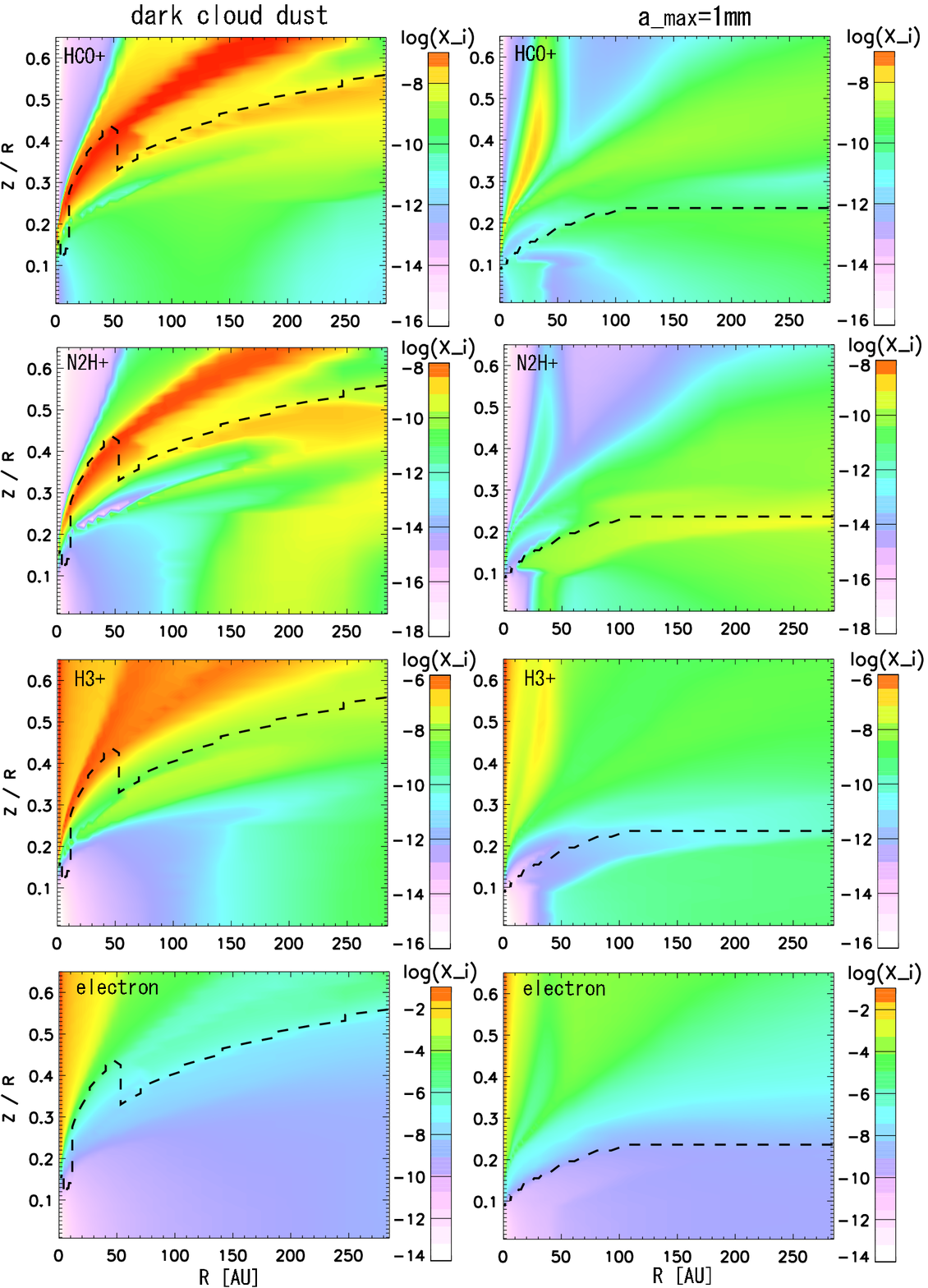}
\caption{2D distributions of HCO$^+$, N$_2$H$^+$, H$_3^+$, and electron calculated using the analytical formulas
in the disk models with dark cloud dust (left panel) and millimeter grains (right panels).
CO and N$_2$ abundances are adopted from the full-network model at $t=1\times 10^5$ yr. The dashed line
depicts the layer above which the electron abundance is adopted from the full network model.
\label{analytic_2D}}
\end{figure}


\begin{figure}
\epsscale{.80}
\plotone{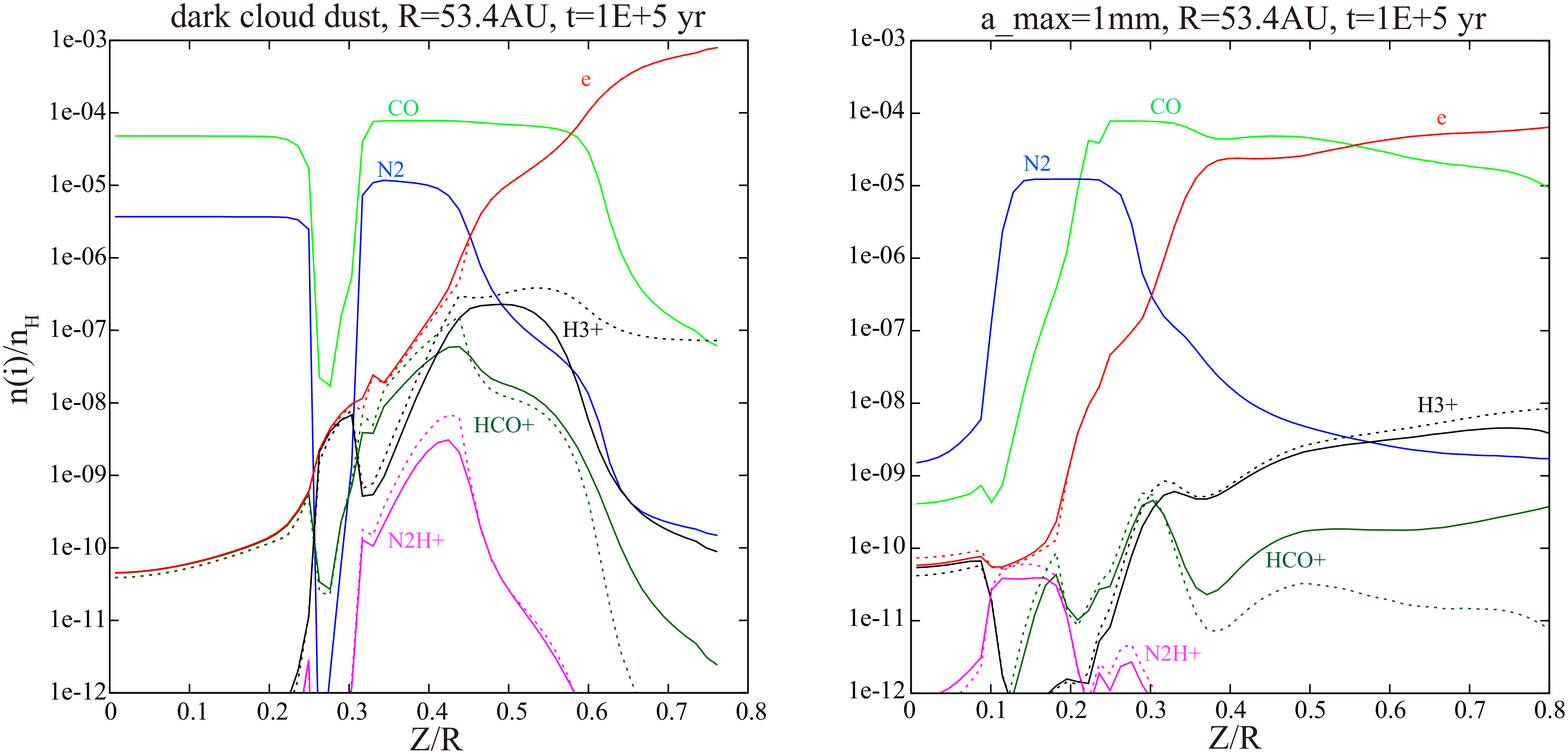}
\caption{Molecular abundances as a function of height from the midplane ($Z/R$) at $R=53.4$ AU
in the disk models with dark cloud dust (left panel) and millimeter grains (right panel) at $t=1\times 10^5$ yr.
The solid lines depict the results of the full-network model, while the dotted lines depict the abundances obtained by the
analytical formulas.
\label{analytic_1D}}
\end{figure}

\begin{figure}
\epsscale{.80}
\plotone{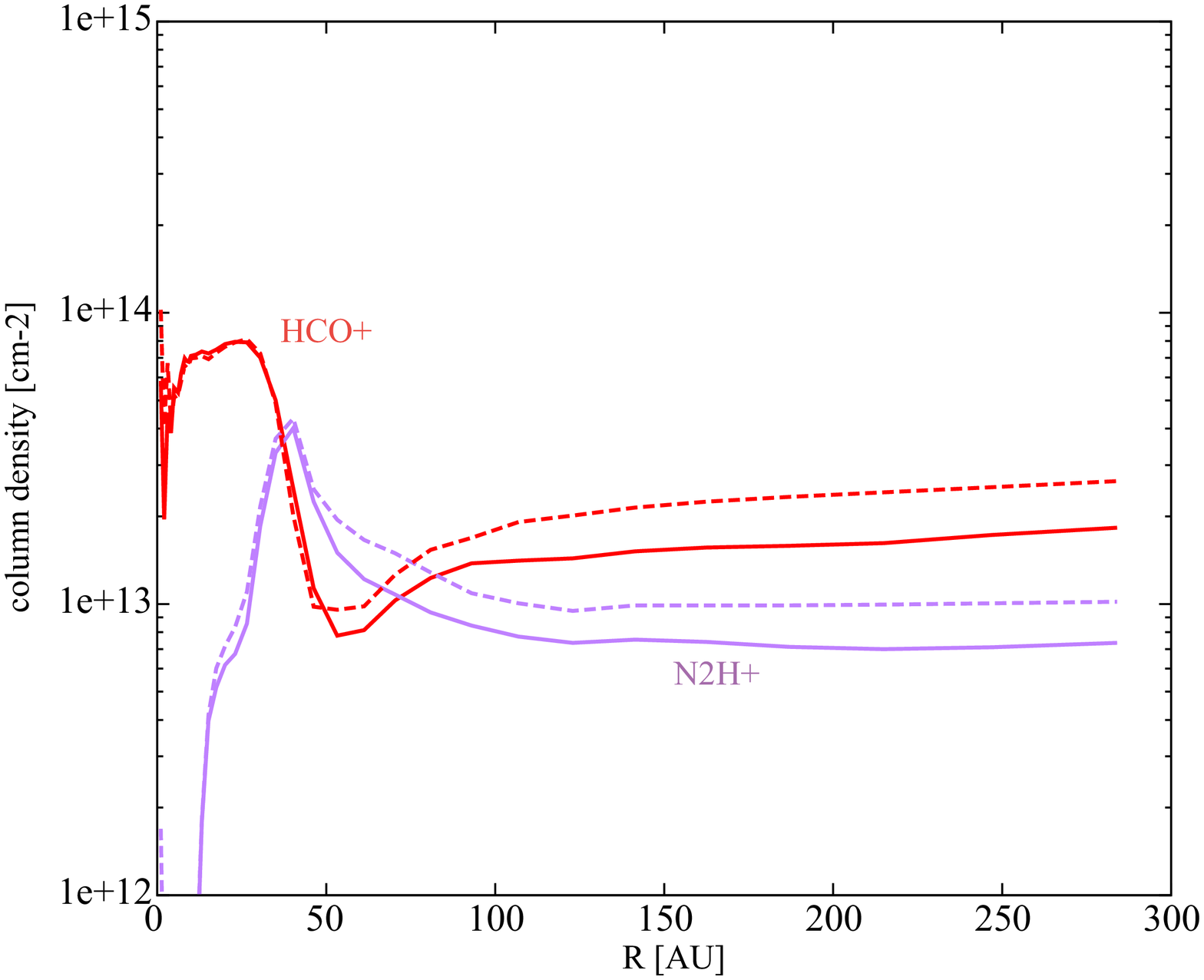}
\caption{Radial distributions of column densities of HCO$^+$ and N$_2$H$^+$ in the full-network model with millimeter grains
at $t=10^5$ yr (solid lines).
The column densities obtained by the analytical formulas are depicted by the dotted lines.
\label{analytic_col}}
\end{figure}

\begin{figure}
\epsscale{1.0}
\plotone{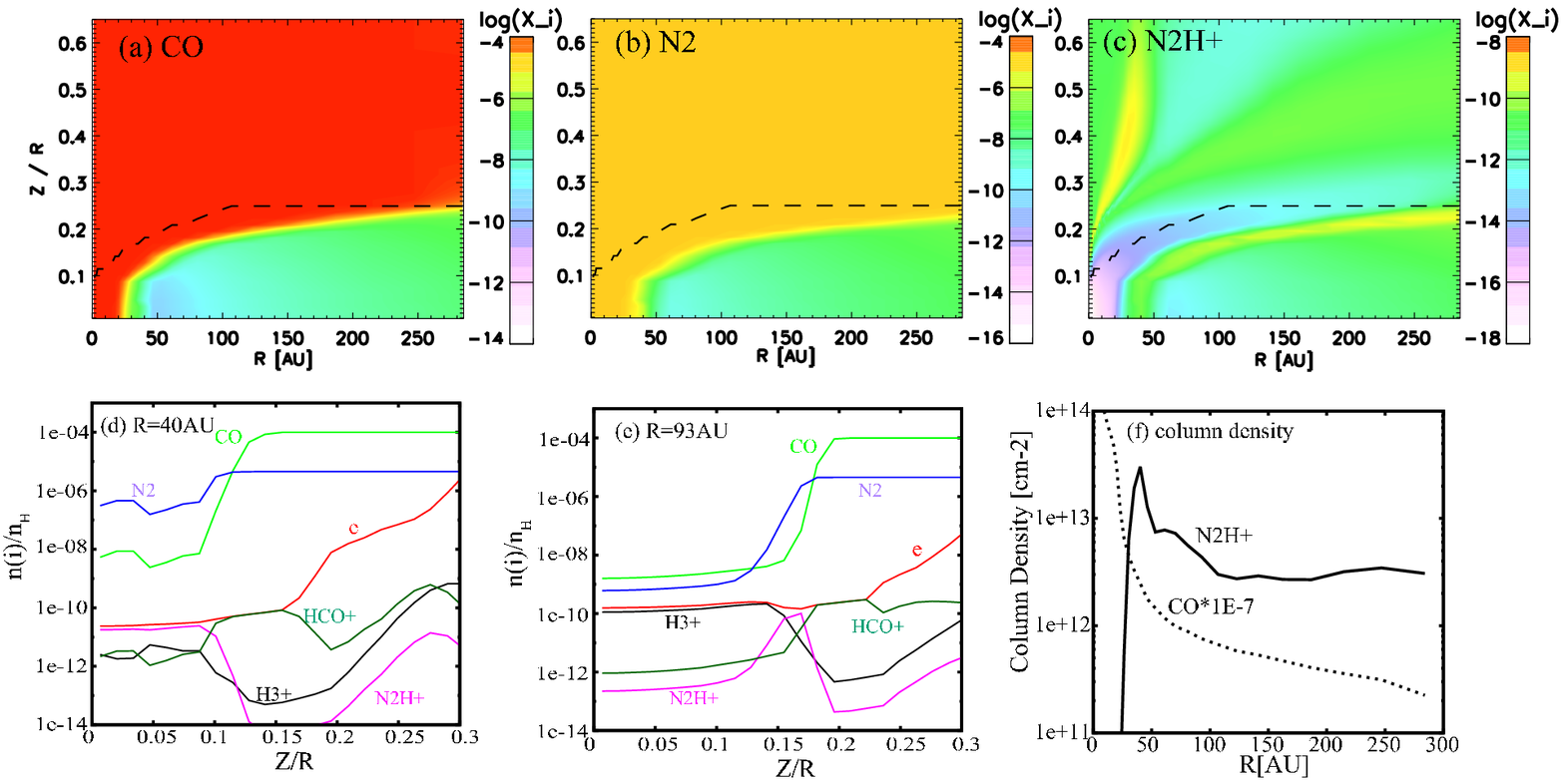}
\caption{(a-c) Distributions of abundances of CO, N$_2$ and N$_2$H$^+$ in the no-sink model.
The desorption energies are set to be $E_{\rm des}$(CO)$=1150$ K and $E_{\rm des}$(N$_2$) $=1000$ K.
The dashed lines depict the layer above which the electron abundance is adopted from
the full-network model.
(d-e) Vertical distributions of molecules at $R=40$ AU and 93 AU.
(f) Column densities of N$_2$H$^+$ and CO. The column density of CO is multiplied by a factor of
$10^{-7}$ to fit in the figure.
\label{analytic_2D_CON2}}
\end{figure}

\begin{figure}
\epsscale{0.8}
\plotone{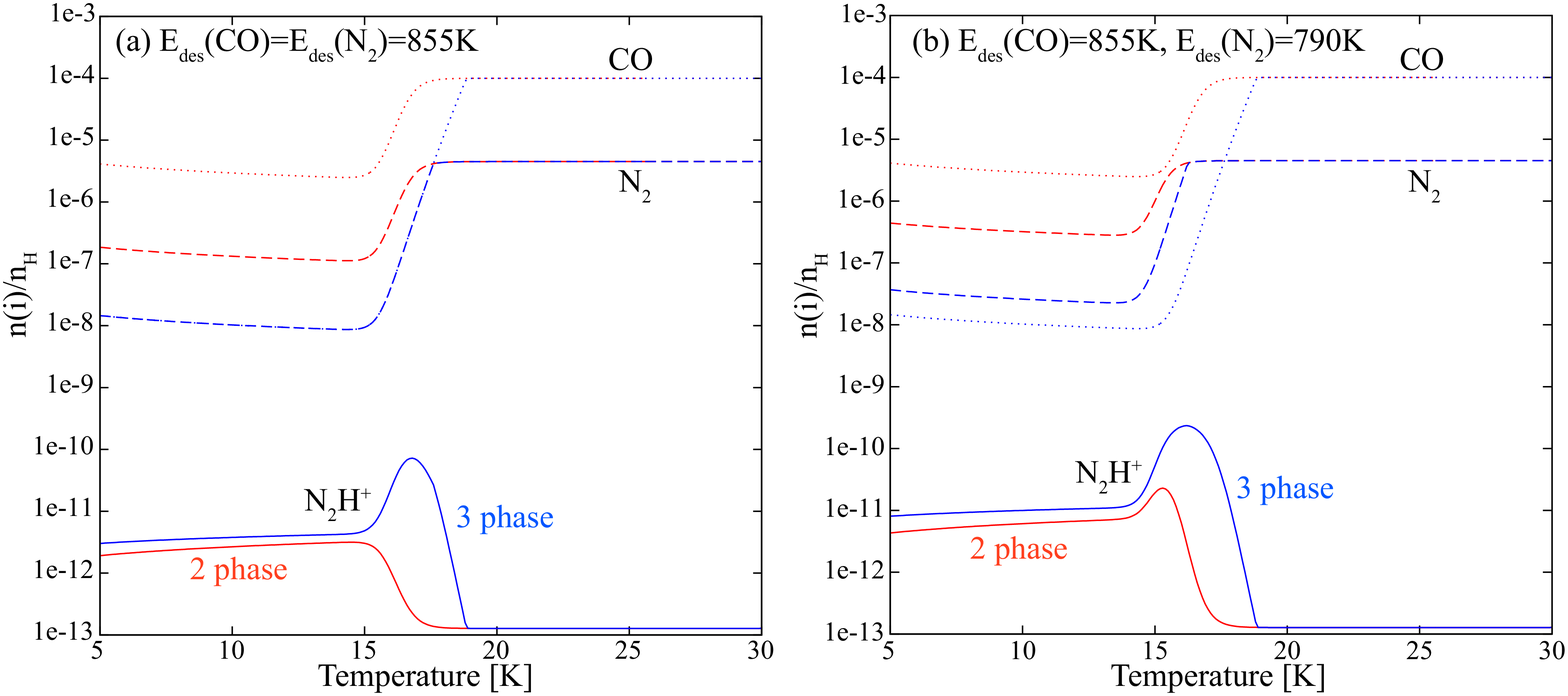}
\caption{Abundances of CO (dotted lines), N$_2$ (dashed lines) and N$_2$H$^+$ (solid lines) as a function of temperature.
The total (gas and ice)
abundances of CO and N$_2$ are assumed to be $1\times 10^{-4}$ and $4.5\times 10^{-6}$, respectively.
The gas density is set to be $1\times 10^8$ cm$^{-3}$ and ionization rate is $5\times 10^{-17}$ s$^{-1}$.
As for the desorption rate, the 2-phase model (eq. \ref{2phase}) is assumed for the red lines,
while the 3-phase model (eq. \ref{eq:3phase}) is assumed for the blue lines.
The desorption energies of CO and N$_2$ are 855 K in panel (a), while $E_{\rm des}$ (CO) is 855 K and  $E_{\rm des}$ (N$_2$) is 790 K in
panel (b).
\label{N2Hp_T}}
\end{figure}

\begin{figure}
\epsscale{1.20}
\plotone{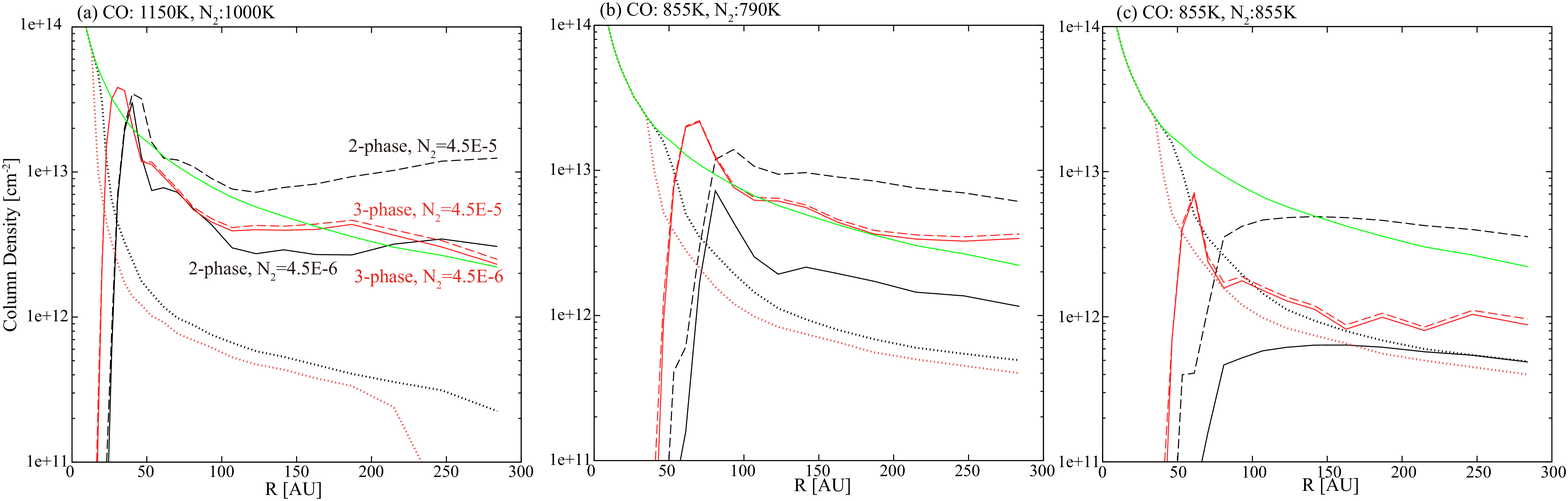}
\caption{Column density of N$_2$H$^+$ in the no-sink models; the total (gas and ice) N$_2$ abundance is set to be
$4.5\times 10^{-6}$ for the solid lines, while it is $4.5\times 10^{-5}$ for the dashed lines.
Black and red lines represent the 2-phase and 3-phase models, respectively.
The dotted lines depict the column density of CO, which is multiplied by a factor of $10^{-7}$ to fit in the figure.
The green lines depict the column density of hydrogen nuclei multiplied by a factor of $10^{-11}$.
N$_2$H$^+$ column density is calculated by the integration
at $Z\le 0.3$ ($R < 50$ AU), $Z\le 0.15$ ($10 < R \le 50$ AU), and $Z\le 0.1$ ($R\le 10$ AU) to avoid the photodissociation region
at the disk surface, while CO and hydrogen column densities are obtained by integrating over the whole disk height.
\label{analytic_col_CON2}}
\end{figure}

\begin{figure}
\epsscale{0.8}
\plotone{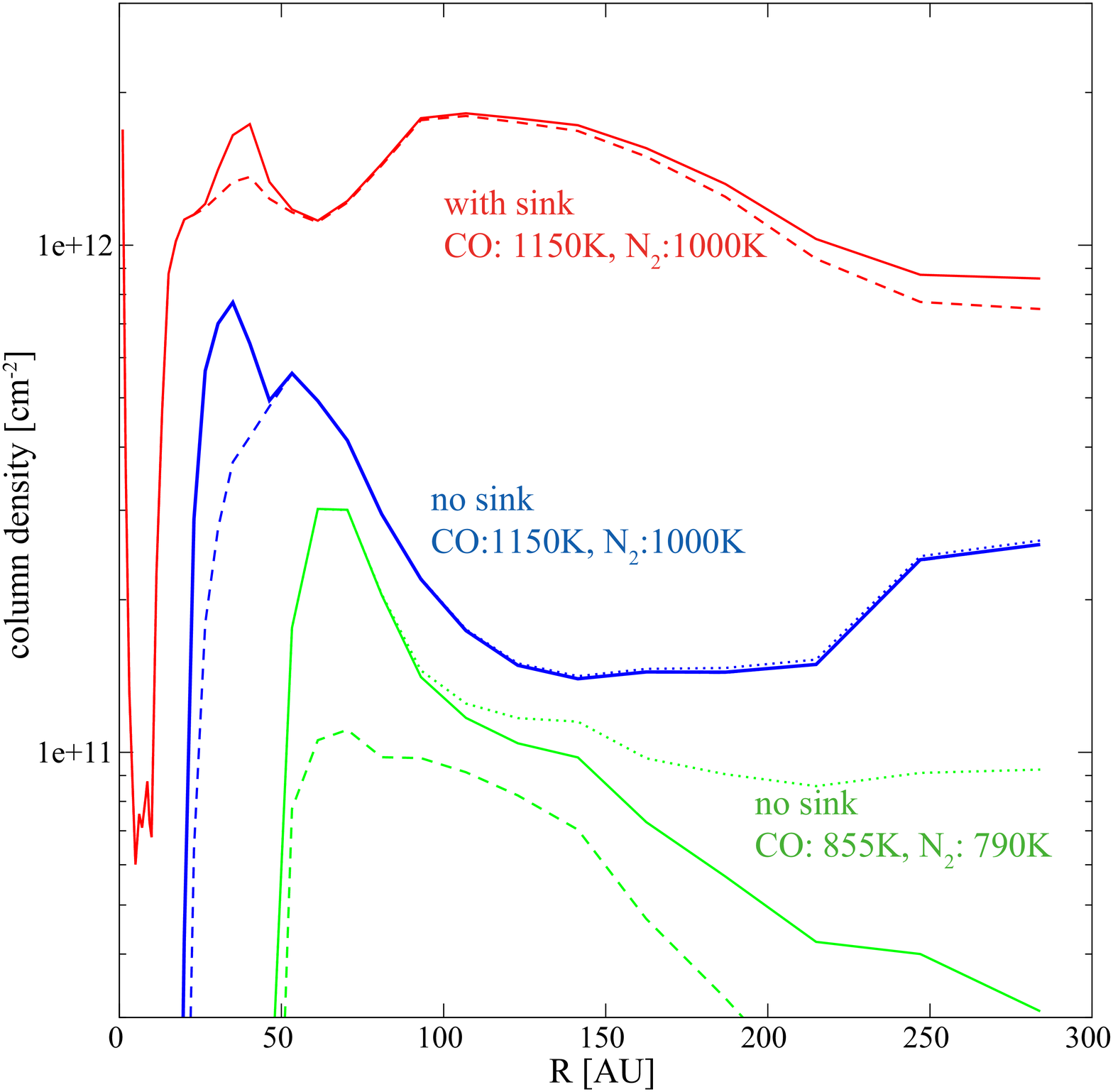}
\caption{Column density of N$_2$H$^+$ in the models without cosmic ray ionization. 
X-ray is the only ionization source for the dashed lines, while the decay of radioactive nuclei is considered for the solid lines
and the dotted lines.
The non-thermal desorption by the cosmic-ray is neglected for the solid lines, while it is included for the dotted lines.
For the blue and green lines, gaseous CO and N$_2$ abundances are given by the 3-phase model with the desorption energies
labeled in the figure. For the red lines, CO and N$_2$ abundances are adopted from the full-network model at $1\times 10^5$ yr,
in which the sink effect on CO is at work.
\label{fig12}}
\end{figure}
\clearpage









\clearpage

\end{document}